\documentclass[%
aps, physrev,
amsmath,amssymb,amsfonts,stmaryrd,
preprint,%
floatfix,nofootinbib]{revtex4-1}

\usepackage{graphicx}
\usepackage{dcolumn}
\usepackage{bm}

\usepackage[utf8]{inputenc}
\usepackage[T1]{fontenc}
\usepackage{mathptmx}
\usepackage{etoolbox}
\usepackage{xcolor}

\makeatletter
\def\@email#1#2{%
	\endgroup
	\patchcmd{\titleblock@produce}
	{\frontmatter@RRAPformat}
	{\frontmatter@RRAPformat{\produce@RRAP{*#1\href{mailto:#2}{#2}}}\frontmatter@RRAPformat}
	{}{}
}%
\newcommand\blfootnote[1]{%
	\begingroup
	\renewcommand\thefootnote{}\footnote{#1}%
	\addtocounter{footnote}{-1}%
	\endgroup
}

\makeatother

\begin{document}

\preprint{APS/123-QED}

\title{\textbf{Simulation of Non-Premixed, Supersonic Combustion using the Discontinuous Galerkin Method on Fully Unstructured Grids} 
}%

\author{Cal J. Rising}
\email{cal.j.rising.civ@us.navy.mil}
\affiliation{Naval Center for Space Technology \\ U.S. Naval Research Laboratory, 4555 Overlook Ave. SW, Washington, DC 20375}

\author{Eric J. Ching}%

\affiliation{Laboratories for Computational Physics and Fluid Dynamics \\ U.S. Naval Research Laboratory, 4555 Overlook Ave. SW, Washington, DC 20375}%

\author{Ryan F. Johnson}
\affiliation{Laboratories for Computational Physics and Fluid Dynamics \\ U.S. Naval Research Laboratory, 4555 Overlook Ave. SW, Washington, DC 20375}%

\date{\today}

\begin{abstract}
In this study, three-dimensional simulations of a reacting hydrogen jet in supersonic crossflow using a structure-preserving discontinuous Galerkin (DG) formulation are examined. The hydrogen jet, with a momentum flux ratio of five, is injected into a high enthalpy crossflow. The sensitivities of the solution to the grid element size and polynomial order are investigated to determine an accurate and computationally efficient approach to simulating high-speed airbreathing propulsion vehicles. The results demonstrate that  DG($p=2$)  solutions, which are nominally third-order accurate in smooth regions of the flow, show reasonable agreement with existing experimental results. The separation shock formation behind the jet is found to be heavily grid dependent and necessary for accurate simulations of the reacting jet in supersonic crossflow. It is determined that the highest considered cell resolution and polynomial order are required to capture the upstream separation shock and consequently the flame stabilization point. The mixing and combustion mode is also determined using the flame index and demonstrates the flow is heavily skewed towards a non-premixed diffusion mode which is consistent with previously run simulations of this case using traditional finite volume schemes and sub grid scale modeling approaches. Beyond this analysis, given the prevalent use of structured hexahedral meshes in the computation of realistic high-speed, chemically reacting flows, the novelty of this work lies in demonstrating accurate simulation of such a flow on a fully unstructured tetrahedral mesh. This highlights the potential of these methods to handle both complicated geometries and complex physics.
\end{abstract}

\maketitle


\section{\label{sec:Intro}Introduction}
\blfootnote{Distribution Statement A. Approved for public release, distribution unlimited.}
Supersonic mixing and combustion are fundamental issues that must be further investigated for reliable implementation of high-speed air-breathing propulsion vehicles. Combustion is challenging in these environments as the need for rapid fuel-air mixing and subsequent burning must occur in the short residence time within the combustor \cite{segal2009scramjet,gruber2004mixing}. Reliable, accurate, and efficient numerical simulation of combustion in these environments is essential for both continued development and deeper fundamental understanding of the governing fluid mechanics. A wide range of fuel–air mixing strategies have been proposed for supersonic combustors in recent years, including struts\cite{genin2010simulation,forster2016analysis}, pylons\cite{doster2009stream,vergine2015supersonic}, cavities\cite{ben2001cavity,gruber2001fundamental}, transverse or angled injection \cite{ben2006time,gamba2015ignition,sharma2020effect}, and hybrid configurations. These varied injection approaches introduce significant complexities into the flowfield, underscoring the need for improved predictive tools and a more comprehensive understanding of the underlying mixing and combustion physics. Transverse fuel injection through a wall orifice is one of the most commonly utilized strategies due to its simplicity in implementation and lack of intrusion into the core flow \cite{huang2016transverse}. Because of the required resolution to accurately model these systems, researchers have primarily focused on investigating and developing different combustion modeling approaches and quantifying their influence on the solution. \cite{candler2017wall,zhao2017large,nilsson2021h2,saghafian2015efficient}


For research and development purposes, Reynolds Averaged Navier Stokes (RANS) simulations that attempt to model all scales in turbulent flows have become the standard \cite{baurle2017hybrid}. However, this modeling approach has limitations in predicting relevant flow fields. These limitations include resolving unsteady physics, accurately predicting shock/boundary layer or shock/fuel jet interactions, and, to some extent, accurately predicting turbulent chemically reacting flows using turbulence closure models \cite{slotnick2014cfd}. Large Eddy Simulations (LES), which resolve large-scale turbulent structures of the flow, are capable of resolving the time-resolved features of a flow while employing an assumption to model the smaller scale turbulence. However, due to the high Reynolds numbers experienced in many relevant flows, the computational expense becomes impractical when trying to resolve the thin viscous layers of wall bounded flows. Hybrid RANS/LES approaches have become more commonly implemented due to their ability to resolve the time-resolved behavior in the combustor while relying on RANS approaches for the wall model \cite{peterson2011simulations,baurle2017hybrid,hassan2017computations,nielsen2021hybrid}. LES/RANS approaches have been used to simulate scramjet facilities and fuel injection configurations and have been shown to provide improved agreement over solely RANS approaches \cite{peterson2011simulations,boles2010large}. However, there can still be difficulties reproducing experimental results when considering challenging regions of the flow (e.g. separated regions, shear layers, reattachment, chemically reacting regions) \cite{peterson2018hybrid,peterson2023simulation}.

In recent years, discontinuous Galerkin (DG) methods have received increased attention in the computational fluid dynamics community as an alternative method for simulating fluid dynamics\cite{gassner2013accuracy,beck2014high,fernandez2017hybridized,lv2023recent,hoskin2024discontinuous}. Some advantages of DG methods compared to traditional finite-volume or finite difference schemes are and the ability to obtain high-quality solutions on unstructured grids for complex geometries, high-order accuracy with a compact stencil, and suitability for modern high performance computing architectures  \cite{ceze2013anisotropic,froehle2014high,arndt2020exadg,cuong2022large}. Additionally, the degree of the solution approximation ($p$) can be increased to improve the resolution on a given grid with elements of size $h$ to limit element count while capturing relevant fluid scales \cite{gassner2013accuracy,beck2014high}. An ongoing area of research in discontinuous Galerkin (DG) methods focuses on retaining their high-order accuracy while addressing broader numerical challenges common across CFD methods, namely, the emergence of nonlinear instabilities in under-resolved regions. These issues are particularly exacerbated by  multicomponent, chemically reacting flows, where sharp gradients and complex coupling increase the difficulty \cite{Chi22,Chi22_2}. To address this issue, a fully conservative, positivity-preserving, and entropy-bounded DG formulation for the chemically reacting Navier-Stokes equations was recently developed~\cite{Chi25}. The formulation reduces spurious pressure oscillations in smooth regions of the flow, ensures nonnegative species concentrations and positive pressure and temperature, and guarantees that the convective contribution discretely satisfies a minimum entropy principle. As such, it can rigorously maintain stability when computing chemically reacting flows involving complex physical phenomena, such as shocks, shear layers, flow separation, and turbulence, even with the low dissipation and high accuracy characteristic of DG schemes.

Existing works have  commonly utilized LES and hybrid RANS/LES approaches where the sub-grid scales (SGS) are resolved using explicit SGS models \cite{candler2017wall,liu2019characteristics,nilsson2021h2,baurle2017hybrid}. An alternative to explicit SGS modeling is implicit LES (ILES), where numerical dissipation due to the discretization is used to represent dissipation associated with the unresolved scales. ILES has been applied to many numerical schemes, including DG methods \cite{beck2014high,fernandez2017hybridized,lv2018underresolved,cuong2022large}. For example, Gassner and Beck investigated the accuracy of DG schemes on Taylor-Green vortex cases with varying Reynolds numbers and found that the ILES DG simulations provided similar results to low order simulations with sub-grid scale models included\cite{gassner2013accuracy}.  Furthermore, even in the context of nonreacting flows, the use of complex SGS models may provide little advantage over simpler SGS models or ILES on a slightly finer grid \cite{spalart2000strategies}. Beyond the pure fluid dynamics, there is additional complexity in proper SGS modeling when considering the interplay between combustion and turbulence\cite{edwards2024multi,pequin2022partially}. In the hybrid RANS/LES simulations performed by Peterson, the solution was found to be sensitive to the turbulence-chemistry interaction (TCI) model, mesh resolution, and chemical reaction model \cite{peterson2023simulation}. These sensitivities make it challenging to properly select a TCI model which will accurately capture the turbulent reacting flow physics. 

Researchers have applied Discontinuous Galerkin (DG) methods to simulate both chemically reacting flows \cite{johnson2020conservative,lv2014discontinuous,lv2017high, gutierrez2022fully, rising2022numerical,Chi25} and external supersonic/hypersonic flows \cite{hoskin2024discontinuous,terrana2020gpu,ching2019shock,bai2022continuous,barter2010shock}. Several works have discussed the benefits of p-adaptivity and p-refinement, however the extension towards high-speed chemically reacting flows remains limited \cite{moxey2017towards,wang2013high,naddei2019comparison,marchal2023extension}. Lv et. al. used an entropy-bounded DG method to explore $h$ and $p$-refinement in compressible decaying isotropic turbulence\cite{lv2018underresolved}. It was shown that $p$-refinement was effective in low to moderate Mach number ranges, while $h$-refinement was more beneficial in supersonic turbulent conditions. {It has also been shown that DG schemes can maintain similar accuracy on tetrahedral and hexahedral meshes, while common finite volume schemes exhibit reduced accuracy on unstructured tetrahedral grids with the same number of degrees of freedom \cite{lv2018underresolved,yoon2007computational,gnoffo2007simulation}. Furthermore, commonly used finite volume approaches for simulating high-speed non-reacting and reacting flows on structured meshes can be sensitive to grid quality, resolution, and alignment with flow features\cite{candler2009current,candler2015next}. While these works have shown promise in leveraging DG on tetrahedral grids as a predictive tool for high-speed propulsion systems, it has yet to be applied in a more application-driven demonstration case. An ideal demonstration would involve key physical complexities—such as shocks, boundary layers, flames, unsteadiness, and geometric intricacies—that benefit from unstructured grids with targeted $h$-refinement with assessment of $p$-refinement.} Fuel injection into a high-speed flow presents a compelling candidate, as it naturally embodies many of these challenges and serves as a relevant, realistic test case.

The current work examines three-dimensional DG simulations of the Stanford University reacting jet in supersonic crossflow (JISCF) experiment \cite{gamba2015ignition} to analyze the effect of $h$ and $p$-refinement on the solution using unstructured tetrahedral grids. The JISCF provides an environment in which the accuracy and cost of the DG method can be investigated on a flow with many relevant supersonic engine features such as boundary layer separation, recirculation zones, shear layers, shock structures, and chemically reacting flow regions \cite{huang2016transverse}. Previous numerical simulations have explored aspects of this particular test case using finite-volume based LES with various subgrid scale and chemistry modeling approaches \cite{liu2019characteristics,nilsson2021h2,candler2017wall,saghafian2015efficient,sharma2024interaction}. {This work expands upon previous research by Rising et. al.~\cite{rising2024use}, which primarily focused on non-reacting mixing statistics on coarse tetrahedral grids using local $h$-refinement and global $p$-refinement. In this study, the reacting flow characteristics  are investigated and quantitative comparisons are made to the available experimental results.} ILES is performed using an extended version of the JENRE\textsuperscript{\textregistered} Multiphysics Framework with the discontinuous Galerkin scheme presented by Ching et. al. \cite{Chi25}. The simulations are performed up to $p=2$ and with four levels of element size refinement to determine the influences of each on the solution accuracy, particularly with respect to the unsteady behavior within the jet shear layers and recirculation zones. The results demonstrate the applicability of using a stable DG method and highlights he promise of computing complex, high-speed, reacting flow on unstructured tetrahedral meshes.

\section{Formulation and methodology}
\subsection{Physical modeling}

The governing equations are the compressible, chemically reacting Navier-Stokes equations, written as
\begin{equation}
\frac{\partial u}{\partial t}+\nabla\cdot\mathcal{F}\left(u,\nabla u\right)-\mathcal{S}\left(u\right)=0,\label{eq:conservation-law-strong-form}
\end{equation}
where $u$ is the vector of state variables, $\nabla u$ is its spatial gradient,
$t$ is time, $\mathcal{F}$ is the flux, and $\mathcal{S}$
is the chemical source term for detailed chemical kinetics~\citep{chemkin89}. The vector of state variables
is expanded as
\begin{equation}
u=\left(\rho v,\rho e_{t},C_{1},\ldots,C_{n_{s}}\right)^{T},\label{eq:reacting-navier-stokes-state}
\end{equation}
where $\rho$ is density, $v=\left(v_{1},v_{2},v_{3}\right)$ is
the velocity vector, $e_{t}$ is the specific total energy, $C_i$
is the molar concentration of the $i$th species, and $n_{s}$ is the number
of species. The partial density of the $i$th species is defined as
\[
\rho_{i}=W_{i}C_{i},
\]
where $W_{i}$ is the molecular weight of the $i$th species. The density is the sum of the partial densities:

\[
\rho=\sum_{i=1}^{n_{s}}\rho_{i}.
\]
The mole and mass fractions of the $i$th species are given
by
\[
X_{i}=\frac{C_{i}}{\sum_{i=1}^{n_{s}}C_{i}},\quad Y_{i}=\frac{\rho_{i}}{\rho}.
\]
The equation of state for the mixture is the ideal gas law, written as
\begin{equation}
P=R^{0}T\sum_{i=1}^{n_{s}}C_{i},\label{eq:EOS}
\end{equation}
where $P$ is the pressure, $T$ is the temperature, and $R^{0}$
is the universal gas constant. The specific total energy is the sum
of the mixture-averaged specific internal energy, $e$, and the specific
kinetic energy, given by

\[
e_{t}=e+\frac{1}{2}\sum_{k=1}^{d}v_{k}v_{k},
\]
where $e$ is the mass-weighted sum of the specific internal
energies of each species,
\[
e=\sum_{i=1}^{n_{s}}Y_{i}e_{i}.
\]
The species internal energies are approximated with temperature-dependent polynomials based
on the NASA coefficients~\citep{Mcb93,Mcb02}.

The flux is the difference between the convective
flux, $\mathcal{F}^{c}$, and the viscous flux, $\mathcal{F}^{v}$,

\[
\mathcal{F}\left(u,\nabla u\right)=\mathcal{F}^{c}\left(u\right)-\mathcal{F}^{v}\left(u,\nabla u\right),
\]
where
\begin{equation}
\mathcal{F}^{c}\left(u\right)=\left(\rho v \otimes v+P\mathbb{I},v\left(\rho e_{t}+P\right),vC_{1},\ldots,vC_{n_{s}}\right)^{T}\label{eq:reacting-navier-stokes-spatial-convective-flux-component}
\end{equation}
 and
\begin{equation}
\mathcal{F}^{v}\left(u,\nabla u\right)=\left(\tau,v \cdot \tau+\sum_{i=1}^{n_{s}}\rho_{i}h_{i}V_{i}-q,C_{1}V_{1},\ldots,C_{n_{s}}V_{n_{s}}\right)^{T}.\label{eq:navier-stokes-viscous-flux-spatial-component}
\end{equation}
$\tau$ is the viscous stress tensor, $q$ is the heat
flux, and $V_{i}$ is the diffusion
velocity of the $i$th species. The source term is written as
\begin{equation}
\mathcal{S}=\left(0,0,\omega_{1},\ldots,\omega_{n_{s}}\right)^{T},
\end{equation}
where $\omega_i$ is the production rate of the $i$th species.

Transport properties are calculated using a mixture-averaged framework. {The $k$th spatial component of the diffusion velocity for the $i$th species is given as}

\begin{equation}
	{\hat{V}_{ik} = \frac{\bar{D}_i}{C_i} \frac{\partial C_i}{\partial x_k} - \frac{\bar{D}_i}{\rho} \frac{\partial \rho}{\partial x_k}}
\end{equation}

{To ensure conservation of mass, i.e. $\sum_{i=1}^{n_{s}}W_iC_iV_{ik} = 0$, a standard correction is applied to the species diffusion velocity\cite{coffee1981transport,houim2011low}} as

\begin{equation}
	{{V}_{ik} = \hat{V}_{ik} - \frac{\sum_{i=1}^{n_{s}}W_i C_i V_{ik}}{\rho}}.
\end{equation}

{The species mixture-averaged diffusion coefficients ($\bar{D}_1,...,\bar{D}_{n_s}$) are determined using the formulation presented in~\cite{Kee89}:}

\begin{equation}
	{\bar{D}_i = \frac{P_{atm}}{P\bar{W}} \frac{\sum_{j=1,j \neq i}^{n_{s}}X_j W_j}{\sum_{j=1,j \neq i}^{n_{s}}X_j/D_{ij}}}
\end{equation}
{where $P_{atm}$ = 101325 Pa, $X_j$ is the mole fraction of species $j$, $D_{ij}$ is the diffusion coefficient of species $i$ to species $j$, and $\bar{W}$ is the mixture molecular weight.} The Wilke mixture model~\citep{Wil50} is employed to calculate dynamic viscosity, while the Mathur model~\citep{Mat67} is used to calculate thermal conductivity. All thermodynamic and transport properties in this work are approximated using differentiable polynomial refits. {The Westbrook hydrogen mechanism as described in Ching et. al is employed ~\cite{WESTBROOK1982191,Chi25}. This mechanism was selected for its previously demonstrated reasonable agreement with experimental results\cite{rising2024use}.} More information can be found in~\cite{johnson2020conservative}.

\subsection{Discretization}

The conservation equations are semi-discretized using the structure-preserving nodal discontinuous Galerkin method described in~\cite{Chi22_2, Chi25}. The formulation reduces spurious pressure oscillations in smooth regions of the flow~\cite{johnson2020conservative} and ensures nonnegative species concentrations, positive pressure and temperature, and bounded specific entropy for the convective contribution to the evolved state. These properties are important for maintaining stability in a mathematically rigorous manner, especially when using low-dissipation, high-order methods on unstructured grids to capture complex physics. The HLLC inviscid flux function~\cite{Tor13} and BR2 viscous flux function~\cite{Bas00} are employed. Strang splitting is applied to decouple temporal integration of the stiff chemical source term from that of the transport operators. {An implicit DG discretization in time~\cite{johnson2020conservative} is used for the former, and third-order strong-stability-preserving Runge-Kutta time integration~\cite{Got01,Ket08} is used for the latter, where the time-step size is governed by
\begin{equation}
	{\mathrm{CFL}=\frac{\Delta t\left(2p+1\right)}{h}\left(\left|v\right|+c\right),}
\end{equation}
with $c$ denoting the speed of sound. Additional Runge-Kutta stages are employed to allow for larger CFL values~\cite{Ket08}}. Artificial viscosity is employed to minimize spurious artifacts at discontinuities. More information on the discretization, nonlinear stability properties, boundary conditions, and artificial-viscosity formulation can be found in~\cite{johnson2020conservative, Chi22_2, Chi25}. The stagnation plenum boundary condition used for the fuel injector in this work is not addressed in the cited references; therefore, it is described in Appendix~\ref{sec:StagnationCondition}. {Several variations of this boundary condition, including that proposed by Rodriguez et al. \cite{rodriguez2018formulation}, were explored. However, the implementation presented in Appendix~\ref{sec:StagnationCondition} proved to be the most robust and stable.}

\begin{figure}[htbp]
		\centering
		\includegraphics[scale = 1.0]{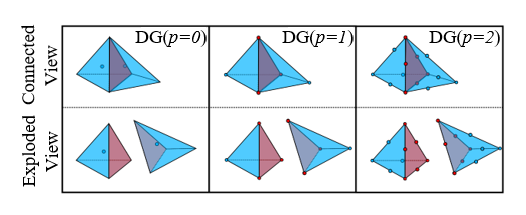}
		\caption{Diagram of tetrahedral cells and their DOF locations.  a) is DG($p=0$),  b) is DG($p=1$),  and c) is DG($p=2$) }
		\label{fig:tets}
\end{figure}

In contrast with finite volume schemes, DG methods feature subcell resolution, with multiple degrees of freedom (DOF) per cell. Figure~\ref{fig:tets} illustrates the DOF locations of two example neighboring tetrahedral cells for  the nodal DG method used in this study. The top row shows the fully connected view, while the bottom row presents an exploded view, where the two triangular cells are separated for visualization purposes. The red triangular face represents the interface between the two cells. Column (a) depicts DG($p=0$) cells, where each cell has a single DOF. This DG($p=0$) formulation is similar to a cell-centered finite volume discretization, with the key difference that finite volume schemes employ additional neighboring cells to construct higher-order approximations. Columns (b) and (c) show DG($p=1$) and DG($p=2$) cells, respectively, where multiple DOFs are contained in each cell. Specifically, each DG($p=1$) cell has four DOFs, and each DG($p=2$) cell has ten DOFs. Consequently, DG methods can often allow for larger cell sizes than finite volume schemes to achieve similar levels of accuracy.

\subsection{Software and Hardware}

The simulations in this work are performed using a modified version of the JENRE\textsuperscript{\textregistered} Multiphysics Framework~\cite{johnson2020conservative} that incorporates the extensions described in~\cite{Chi22_2, Chi25}. The coarser cases are run on eight NVIDIA A100 GPUs on a local cluster at the Laboratories for Computational Physics and Fluid Dynamics, while the finest cases use 32 NVIDIA V100s on the Lawrence Livermore National Laboratory's High Performance Computing system Lassen~\cite{llnl_lassen}. { Simulations were performed using DG($p=0$), DG($p=1$), and DG($p=2$) schemes, where $p+1$ is the nominal order of accuracy for smooth flows. While higher polynomial degrees can improve accuracy per degree of freedom~\cite{wang2013high, lv2018underresolved}, they may also introduce a memory burden, especially on GPUs with multiple species, which would require further code optimization to address beyond the scope of this study.}

\section{Problem Description}
\begin{figure}[htbp]
	\centering
	\includegraphics[scale = 0.53]{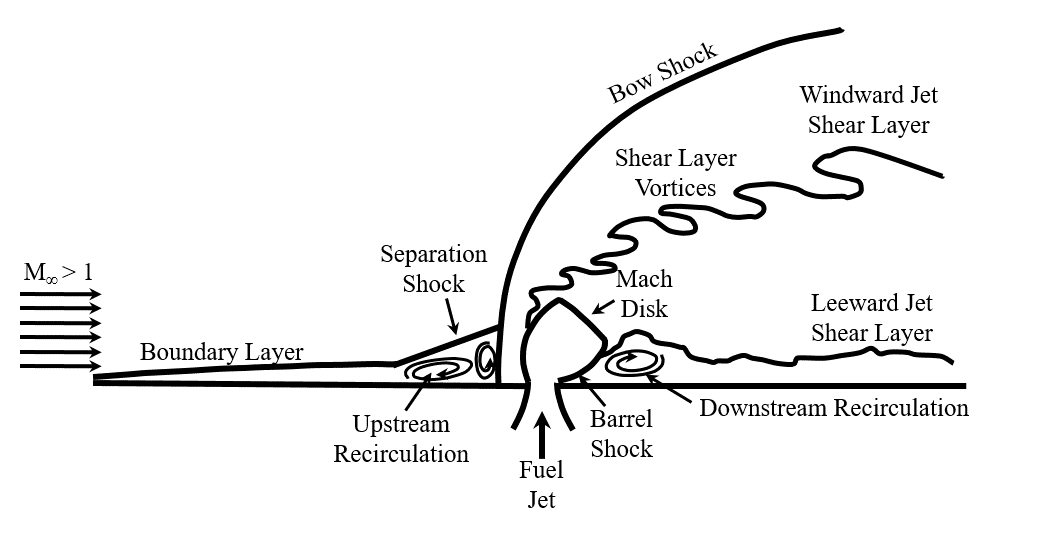}
	\caption{Centerline schematic diagram of primary flow features present in a transverse jet in supersonic crossflow}
	\label{fig:Schematic}
\end{figure}

The jet in supersonic crossflow (JISCF) generates a complex, interconnected system of vortical structures and shocks \cite{fric1994vortical}. Figure \ref{fig:Schematic} depicts a schematic of the primary flow features along the centerline.  The fuel jet obstructs the primary crossflow, forming a three-dimensional bow shock upstream of the injector. This bow shock increases wall pressure, causing boundary layer separation and an upstream recirculation region.  The characteristics of this separated region depend strongly on the incoming boundary layer as a turbulent boundary layer can result in larger separation larger separation region \cite{dowdy1963investigation}. An oblique shock forms from the separation region and intersects the bow shock. This separation region is required to resolve as it serves as a local upstream region where fuel mixing and flame stabilization can occur. This recirculation region also serves to redistribute fuel along the plate as it gets redirected around the jet via the horseshoe vortex system. Additionally, the high temperature and pressure region at the bow shock also creates a local region conducive to autoignition of the fuel. 

An underexpanded jet injected into a crossflow forms a characteristic barrel shock.  The interaction between the jet and the crossflow generates a windward shear layer, prone to Kelvin-Helmholtz instabilities and subsequent vortex formation.  Downstream of the barrel shock, a leeward shear layer develops alongside a recirculation region caused by flow blockage. These shear layers, along with a counter-rotating vortex pair (CVP) oriented in the streamwise direction along the jet (not shown), are critical for fuel-air mixing and enhanced combustion.  Resolving these flow features is essential for accurately predicting flame stabilization and combustion characteristics observed experimentally.

\subsection{Geometry and Boundary Conditions}

\begin{figure}[htbp]
		\centering
		\includegraphics[scale = 0.43]{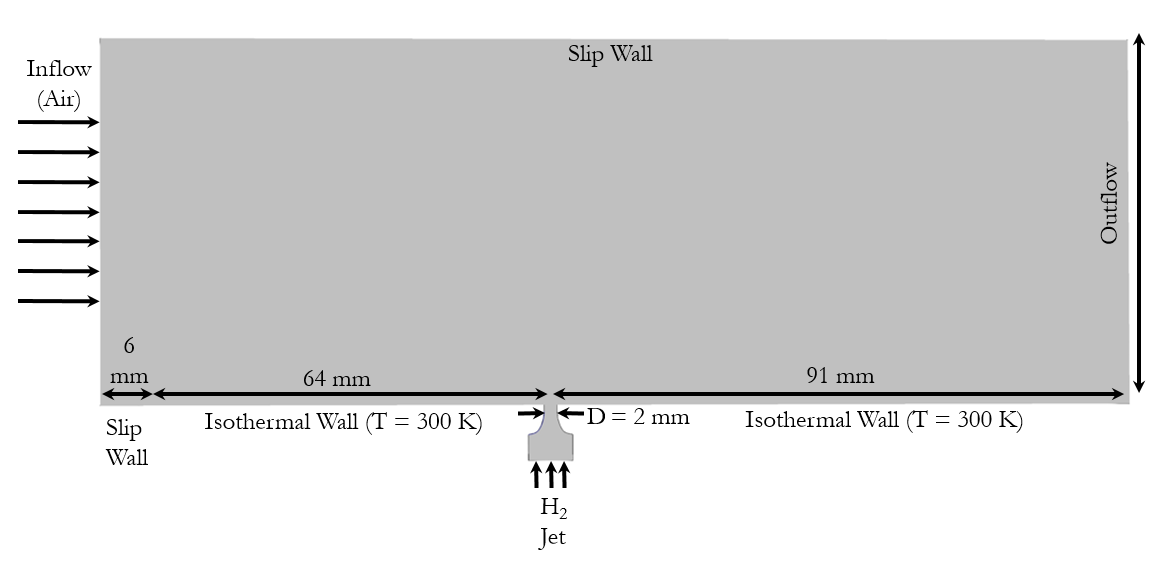}
		\caption{Computational domain and boundary conditions for reacting jet in supersonic crossflow}
		\label{fig:geo}
\end{figure}

{This study simulates the reacting jet in supersonic crossflow experiments conducted at the Expansion Tube Facility of the High Temperature Gas Dynamics Laboratory at Stanford University \cite{gamba2015ignition}. This work builds upon simulation capabilities developed in previous research by Rising et al., utilizing the same geometry and boundary conditions \cite{rising2024use}. The geometry and boundary conditions along the centerplane of the plate are presented in Figure \ref{fig:geo}. The computational domain is a flat plate, 155 mm wide and 100 mm long, with hydrogen injected into the crossflow from a 2 mm diameter orifice located 64 mm downstream of the leading edge.  An isothermal boundary condition of 300 K is applied to the flat plate walls, due to the short experimental run times.  Slip wall boundary conditions are applied to both the side and top walls.}

{The injector boundary conditions are determined based on a jet-to-crossflow momentum flux ratio ($J$), which is defined as}

\begin{equation}
		J = \frac{\rho_j \boldsymbol{v_j}^2}{\rho_{\infty}\boldsymbol{v_{\infty}}^2} = \frac{P_j\gamma_j \mathrm{Ma}_j^2}{P_{\infty}\gamma_{\infty} \mathrm{Ma}_{\infty}^2}
\end{equation}
{\noindent where Ma is the Mach number and $\gamma$ is the ratio of specific heats. This study will focus experimental case from Gamba and Mungal\cite{gamba2015ignition} which was run at momentum flux ratio of J=5. Table \ref{table:Simulation Test Conditions} presents the crossflow boundary conditions, injector stagnation conditions, and species mass fractions ($Y_i$) used in the simulations. The subscripts $\infty$ and $t$ denote the crossflow conditions and injector stagnation conditions, respectively.} 

\begin{table}[htbp]
		\begin{center}
			\begin{tabular}{|c|c|c|c|}
				\hline
				\hline
				\multicolumn{2}{|c|}{Crossflow} &
				\multicolumn{2}{|c|}{Jet} \\
				\hline
				\hline
				$\mathrm{Ma}_{\infty}$ & 2.48 & $\mathrm{Ma}$ & 1.0 \\
				\hline
				$P_{\infty}$ (kPa) & 40 & $P_t$ (kPa) & 2034 \\
				\hline
				$T_{\infty}$ (K) & 1400 & $T_t$ (K) & 300 \\
				\hline
				$Y_{N_2}$ & 0.7686 & $Y_{N_2}$ & 0\\
				\hline
				$Y_{O_2}$ & 0.2314 & $Y_{O_2}$ & 0\\
				\hline
				$Y_{H_2}$ & 0 & $Y_{H_2}$ & 1\\
				\hline
				\hline
			\end{tabular}
			\caption{Crossflow and transverse jet flow conditions}
			\label{table:Simulation Test Conditions}

		\end{center}
\end{table}

\begin{figure}[htbp]
	\centering
	\includegraphics[scale = 1.0]{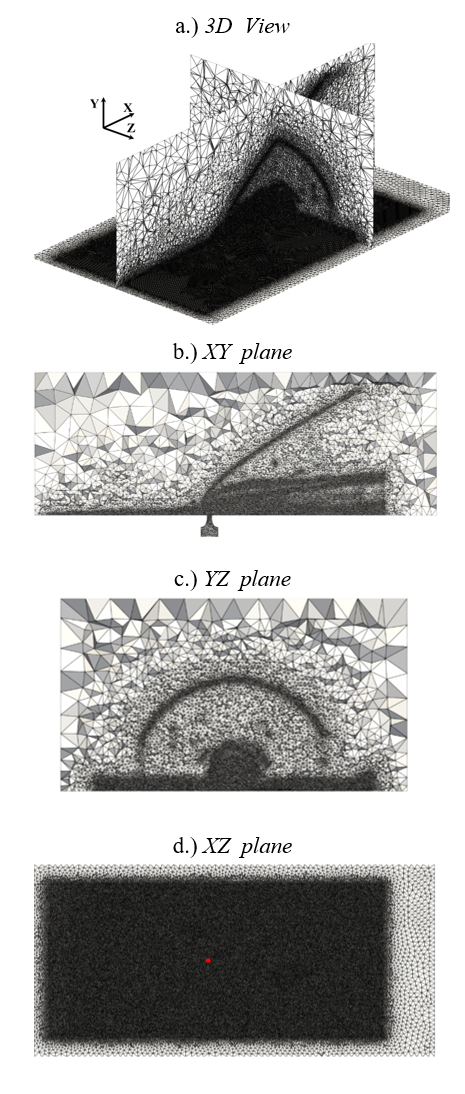}
	\caption{{a.) 3D view, b.) XY Centerplane view, and c.) YZ plane at X/D = 15 view and d.) XZ plane view at Y/D = 0.25 of 300 $\mathrm{{\mu}m}$ AMR refined mesh}}
	\label{fig:mesh}
\end{figure}

\subsection{Mesh Generation}
An unstructured tetrahedral mesh using Adaptive Mesh Refinement (AMR) on RANS simulations is developed using the commercial software STAR-CCM+ \cite{ccm0}. The RANS simulations capture the average flow field and are able to capture the extents of fuel injection and the key shock structures present in this configuration. AMR was performed using the Laplacian of Mach number and the mass fraction of hydroxyl ($Y_{OH}$) as refinement indicators. Within the refinement regions, target element sizes of 600, 450, 300, and 250 $\mathrm{{\mu}m}$ were used to capture relevant flow features. Elsewhere, element sizes are allowed to grow up to a maximum of 1000 $\mathrm{{\mu}m}$.  Boundary layer refinement sensitivity was assessed by further refining wall element size to 150 $\mathrm{{\mu}m}$ (for the 250 $\mathrm{{\mu}m}$ case) and 75 $\mathrm{{\mu}m}$ (for the 450 $\mathrm{{\mu}m}$ case). An example of a refined mesh is shown in Figure \ref{fig:mesh}. {The local refinement along the bow shock, jet shear layer, and boundary layer region can be seen in both Figures \ref{fig:mesh}b and \ref{fig:mesh}c. The mesh refinement in the near wall region is highlighted in Figure \ref{fig:mesh}d, as shown by the constant element sizing within Z/D = $\mathrm{\pm}$ 20. These demonstrate the ability of the meshing approach to capture the key fluid dynamic features presented in Figure \ref{fig:Schematic}.}

Previous research by Moura et. al. \cite{moura2017eddy} has shown the effective resolution in DG calculations can be approximated as $h/(p+1)$. The effective resolution of the simulations using different $p$ and $h$ values, as well as the total element count for each mesh, is presented in Table \ref{table:Mesh Conditions}. It should be noted that given the same effective resolution, polynomial refinement has been shown to improve solution quality to a greater extent than mesh refinement in the context of strong bow shocks~\cite{Chi19} and compressible turbulence (unless the turbulence Mach number is very high)~\cite{lv2018underresolved}. Prior finite volume simulations of this configuration employed hexahedral element counts ranging from 11 to 295 million\cite{liu2019characteristics,nilsson2021h2,candler2017wall,saghafian2015efficient,sharma2024interaction}.  Direct comparison of grid resolution requirements is difficult due to variations in geometry, sub-grid scale models, and chemistry approaches. However, the degrees of freedom (DOFs) for each grid in Table II are comparable to previous studies despite using coarser tetrahedral elements. For example, the finest grid using 20.5x10\textsuperscript{6} elements has 205x10\textsuperscript{6} p=2 DOFs, which is the same order of magnitude of the finer resolution FVM studies. 

\begin{table}[htbp]
	\begin{center}
		\small
		\begin{tabular}{|c|c|c|c|c|c|}
			\hline
			\hline
			Jet/Wall Size & DG($p=0$) & DG($p=1$) &DG($p=2$) & Elements \\
			\hline
			\hline
			&&&&\\[-1em]
			600/600 $\mathrm{{\mu}m}$ & 600 & 300 & 200 & 1.8x10\textsuperscript{6} \\
			\hline
			&&&&\\[-1em]
			300/300 $\mathrm{{\mu}m}$ & 300 & 150 & 100 &  8.8x10\textsuperscript{6} \\
			\hline
			&&&&\\[-1em]
			250/150 $\mathrm{{\mu}m}$ & 250/150 & 125/75 & 83.33/50 & 13.3x10$^6$ \\
			\hline
			&&&&\\[-1em]
			450/75 $\mathrm{{\mu}m}$ & 450/75 & 225/37.5 & 150/25 & 20.5x10$^6$ \\
			\hline
		\end{tabular}
		\caption{Effective resolution for different element sizes ($h$) and polynomial orders ($p$)}
		\label{table:Mesh Conditions}
	\end{center}
\end{table}

\section{Results}

\begin{figure}[htbp]
	\centering
	\includegraphics[scale = 0.96]{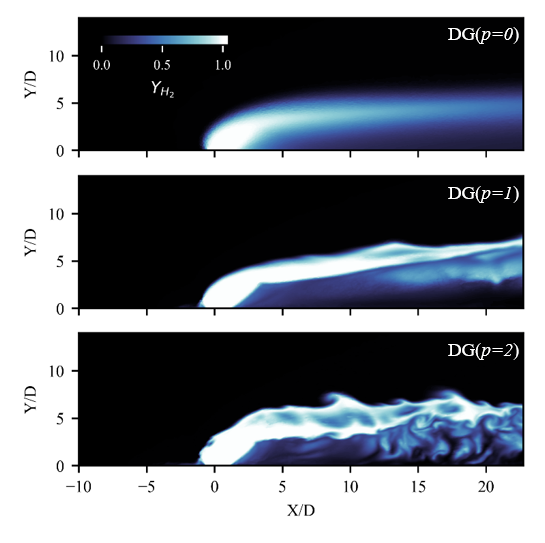}
	\caption{Centerline contour of $Y_{H_2}$ mass fraction on the 300 $\mathrm{{\mu}m}$ grid for each polynomial level}
	\label{fig:H2}
\end{figure}

\subsection{Flow Field Characteristics}

\begin{figure*}[htbp]
	\centering
	\includegraphics[scale = 0.99]{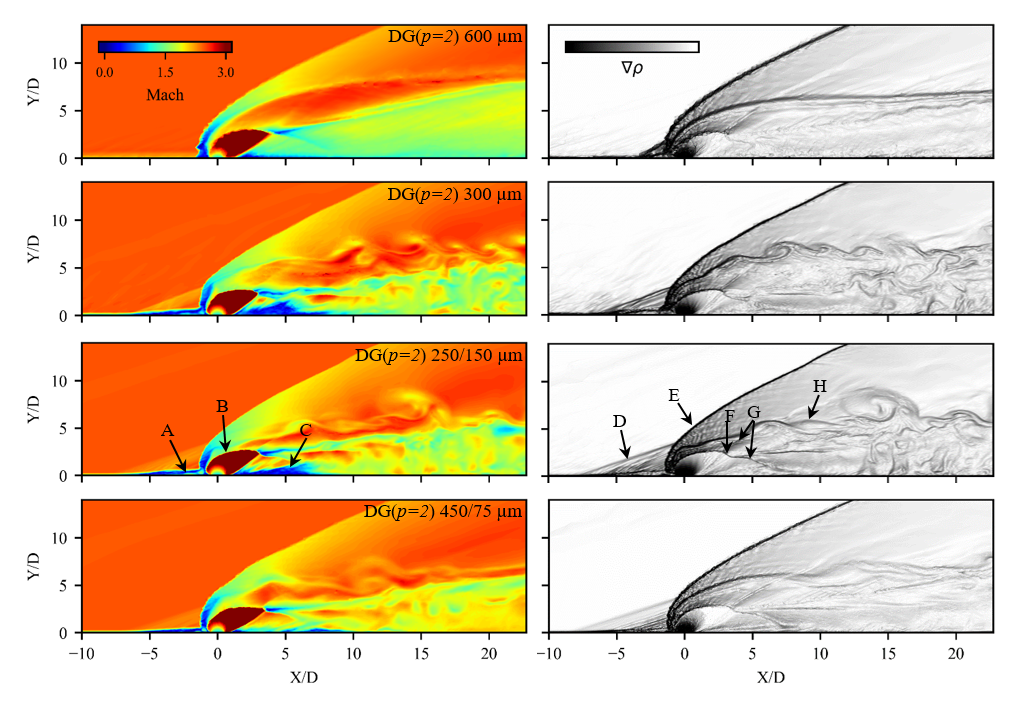}
	\caption{Instantaneous contours of Mach number and numerical schlieren for DG($p=2$) solution on grids with 600, 300, 150 $\mathrm{\mu}$m wall elements. The labels represent: A - upstream recirculation zone, B - barrel shock region, C - downstream recirculation zone, D - separation shock, E - bow shock, F - Mach disk, G - Mach disk reflected shocks, H - windward jet shear layer.}
	\label{fig:M_NS}
\end{figure*}

Instantaneous hydrogen mass fraction contours, shown in Figure \ref{fig:H2} for the 300 $\mathrm{\mu}$m grid at each polynomial order, are used to analyze the jet shear layer and mixing characteristics. The contours highlight the influence of global p-refinement on the shear layer structures. The 300 $\mathrm{\mu}$m grid was selected for analysis of polynomial order effects, as it represents the minimum resolution necessary for some shear layer features to be resolved. The DG solution with polynomial order $p=0$ lacks sufficient resolution, resulting in a diffuse hydrogen distribution that fails to accurately capture the hydrogen jet. As order of accuracy of the polynomial is increased to DG($p=1$), the jet begins to take a more defined shape, with some fluid dynamic driven instabilities. In the DG($p=2$) solution, the KH instability begins to appear more significantly in the windward jet shear layer. Furthermore, increased hydrogen entrainment towards the plate is observed, along with the development of additional structures in the leeward shear layer. Similar results hold for the other more refined grids described in Table \ref{table:Mesh Conditions},  highlighting that third order-accuracy elements are required. The results in the following section will focus on the DG($p=2$) simulations; however, DG($p=1$) results will be included to highlight the influence of the polynomial order on key statistics.

To further examine the jet shear layer and shock structures, instantaneous contours of Mach number and numerical schlieren are presented in Figure \ref{fig:M_NS}. Across all grid resolutions, the barrel shock and Mach disk remains constant in size and location. When comparing the windward shear layer, the 600 $\mathrm{\mu}$m simulation exhibits minimal vortex structures. In contrast, the 300 and 250/150 $\mathrm{\mu}$m meshes lead to more prominent vortex structures as the jet shear layer turns and separates from the bow shock. Additionally, there are noticeably more turbulent structures that appear downstream of the jet in the near-wall region. The most significant difference between each grid is the location of the upstream separation shock caused by the fuel jet. The separation shock moves upstream from X/D $\approx$ -2 to X/D $\approx$ -8 as the near wall element size decreases. This is likely a result of the increased resolution in the boundary layer, which encounters the transverse fuel jet. Both the 250/150 and 450/75 $\mathrm{\mu}$m grids provide separation shock locations near X/D $\approx$ -8, which agrees with experimental images of Gamba and Mungal in the $J=5$ case \cite{gamba2015ignition}. Although the 450/75 $\mathrm{\mu}$m grid improves resolution in the boundary layer, the coarser resolution in the shear layer region limits the development of turbulent structures. 

\begin{figure}[htbp]
	\centering
	\includegraphics[scale = 0.98]{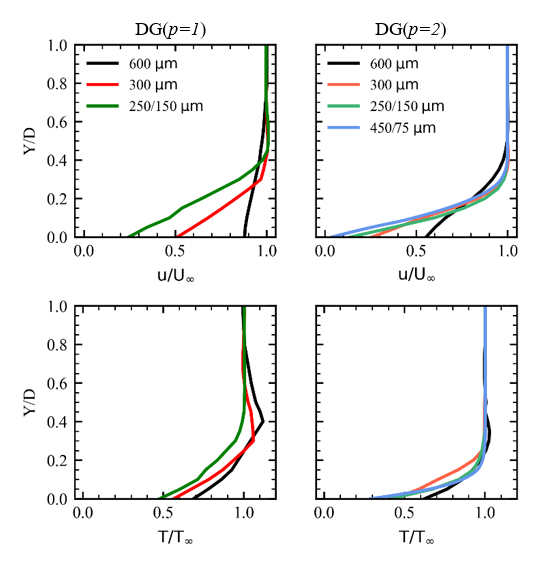}
	\caption{Average velocity profiles for each mesh size and polynomial level extracted X/D = -12.5}
	\label{fig:BL_Fig}
\end{figure}
Average velocity and temperature profiles of the incoming laminar boundary layer are
presented in Figure \ref{fig:BL_Fig} to analyze the effects of the nearwall grid resolution. The velocity and temperature profiles are extracted at X/D = -12.5 for all grids to ensure that the profiles are not influenced by the separation shock and recirculation region. It is seen that all DG($p=1$) solutions poorly capture the  boundary layer profile, further demonstrating the need for third order accurate DG($p=2$) elements. As the resolution is improved the boundary layer becomes more resolved in the near wall region with both the 450/75 and 250/150 $\mathrm{\mu}$m grids providing the best results. All grids aside from the 600 $\mathrm{\mu}$m predict a boundary layer height of approximately 0.4 Y/D (or 0.8 mm), which is consistent with the laminar boundary layer thickness predictions made by Gamba et. al.\cite{gamba2015ignition}. Similarly, the temperature profiles exhibit behavior consistent with the velocity profiles, demonstrating that the 450/75 $\mathrm{\mu}$m and 250/150 $\mathrm{\mu}$m grids provide the anticipated results for both. These predictions demonstrate a correlation between near-wall resolution and boundary layer profile accuracy, and the accurate prediction of both separation shock location and upstream recirculation region behavior.

\begin{figure}[htbp]
	\centering
	\includegraphics[scale = 0.96]{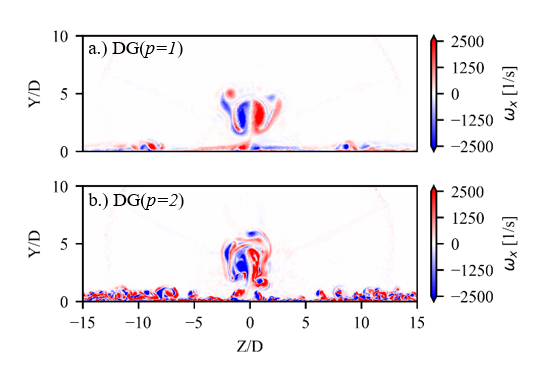}
	\caption{{Instantaneous streamwise vorticity ($\omega_x$) at at X/D = 15 for a.) DG($p=1$)  and b.) DG($p=2$) for the 250/150 $\mathrm{\mu}$m grid}}
	\label{fig:stream_wise}
\end{figure}

{The effect of polynomial order on jet mixing is further examined through a comparison of instantaneous streamwise vorticity ($\omega_x$) contours.} Figure \ref{fig:stream_wise} presents these visualizations for DG(p=1) and DG(p=2) solutions on the 250/150 $\mathrm{\mu}$m grid. It is seen that the DG($p=1$) solution is capable of resolving the features away from the wall as the streamlines depict the CVP formation and bow shock location. When the resolution is improved with DG($p=2$) elements, the bow shock and CVP are similar. A key discrepancy is that the two vortex lobes formed are less diffuse and narrower in the DG($p=2$) solution. The primary difference between the solutions is the turbulence content captured within the boundary, as depicted by the vortex production within the boundary layer region. This is a further indication of the ability of increased polynomial representation to capture the relevant flow physics on underresolved grids.

\subsection{Combustion Characteristics}

\begin{figure}[htbp]
	\centering
	\includegraphics[scale = 0.95]{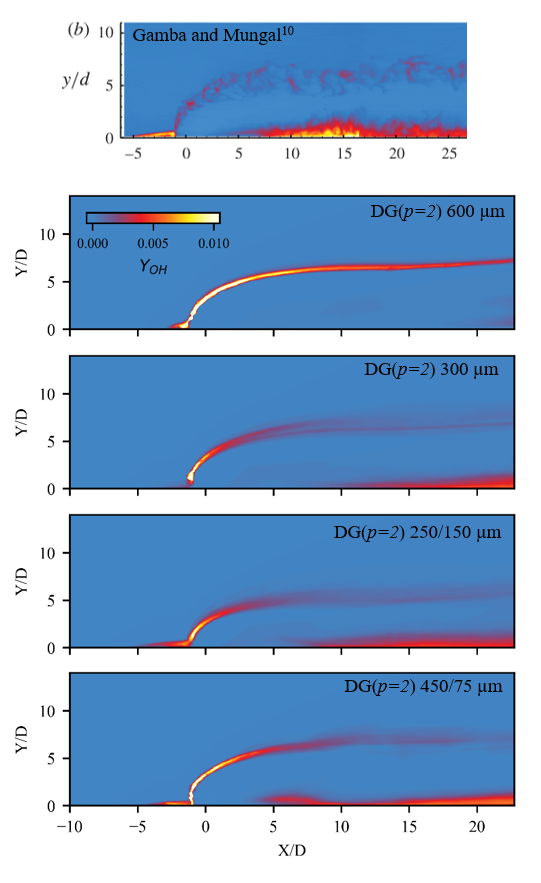}
	\caption{{Experimental centerline average OH-PLIF image from Gamba and Mungal\cite{gamba2015ignition} and average centerline contours of $Y_{OH}$ mass fraction for DG($p=2$) solution on grids with $h=600$, 300, 250/150, and 450/75 $\mathrm{\mu}$m  at the wall.}}
	\label{fig:oh_mean}
\end{figure}

\begin{figure}[t]
		\centering
		\includegraphics[scale = 0.95]{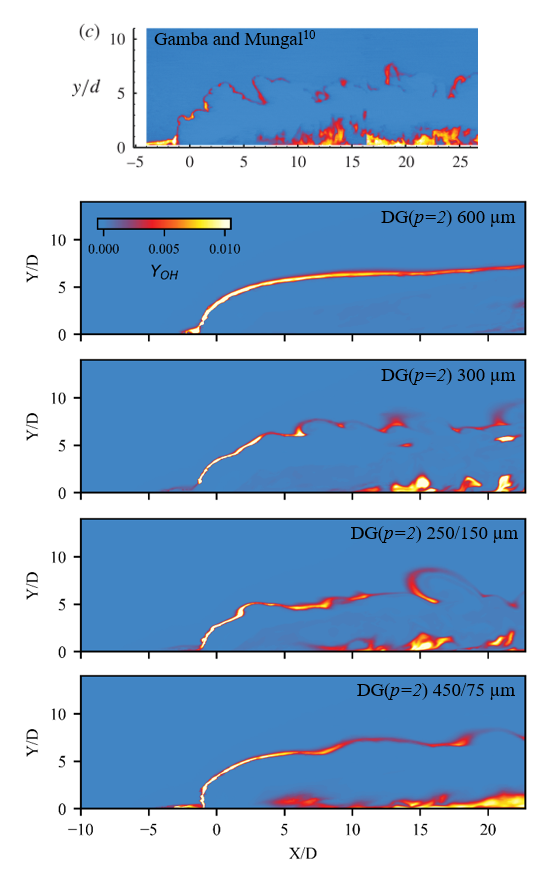}
		\caption{{Experimental centerline instantaneous OH-PLIF image from Gamba and Mungal\cite{gamba2015ignition} and instantaneous centerline contours of $Y_{OH}$ mass fraction for DG($p=2$) solution on grids with $h=600$, 300, 250/150, and 450/75 $\mathrm{\mu}$m  at the wall.}}
		\label{fig:oh_inst}
\end{figure}

Having established the ability to accurately capture the relevant fluid dynamics,  the reacting flow characteristics are examined in more detail. Figure \ref{fig:oh_mean} and \ref{fig:oh_inst} present time averaged and instantaneous contours of OH mass fraction ($Y_{OH}$) are shown to identify key features such as the flame stabilization points, reacting shear layer location and thickness, and comparison to experimental OH Planar Laser Induced Fluorescence (PLIF) measurements. The solution on the 600 $\mathrm{\mu}$m grid demonstrates consistent combustion along the windward jet shear layer. In the further refined grids, OH production occurs further upstream of the jet and closer to the jet in the downstream region. Additionally, along the windward jet shear layer, there is intermittent blowout, represented by discontinuous OH production, which is more consistent with the experimental measurements. A difference between the 300 and 150 $\mathrm{\mu}$m grids is the OH layer penetrates higher to Y/D = 8 on the 300 $\mathrm{\mu}$m grid and approximately Y/D = 6 on the 150 $\mathrm{\mu}$m grid. This is also attributed to the improved resolution in the boundary layer, which allows additional vorticity to be transferred to the windward jet shear layer, as discussed by Pizzaia et. al. \cite{pizzaia2018effect}.

\begin{figure}[t]
		\centering
		\includegraphics[scale = 1.0]{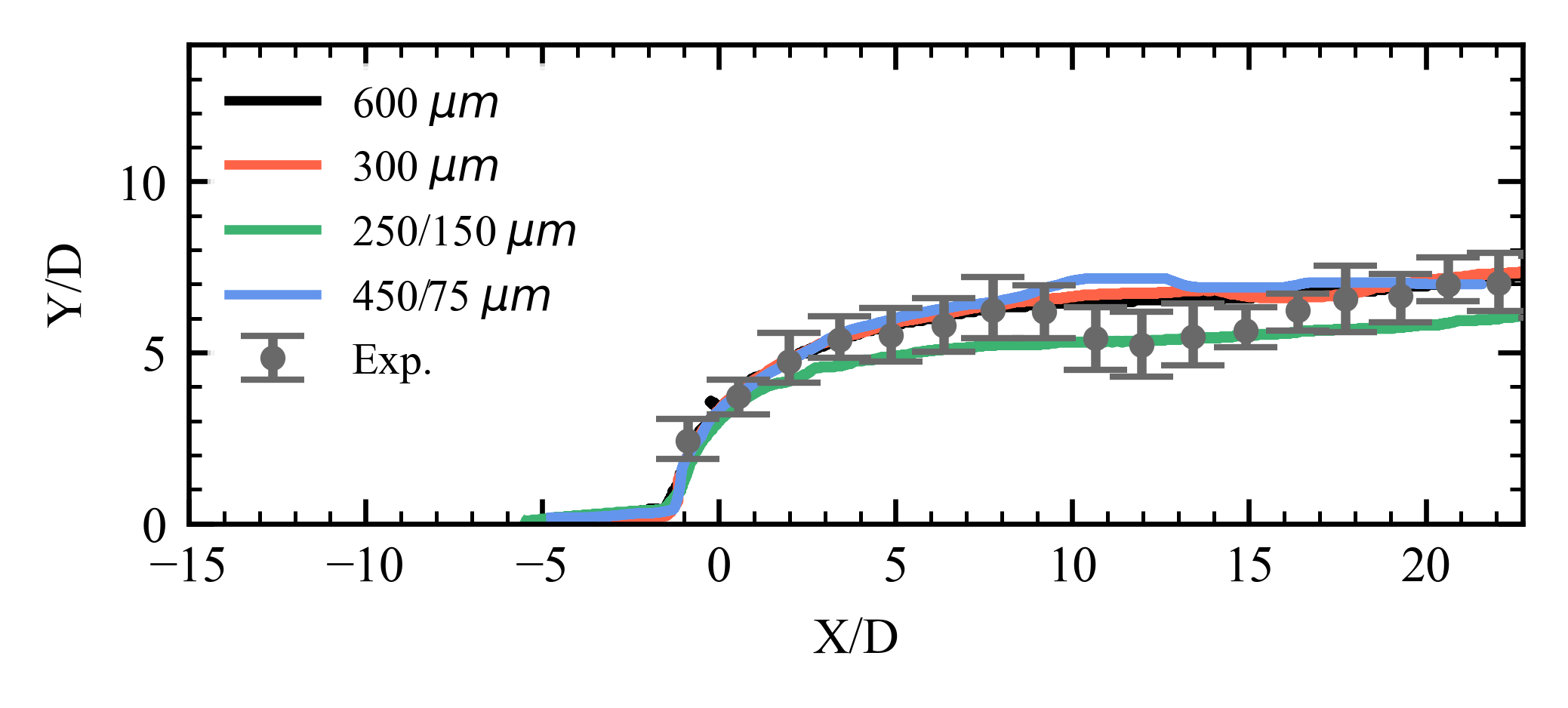}
		\caption{Mean OH layer location for DG($p=2$) simulations on each grid size}
		\label{fig:oh_loc}
\end{figure}

To quantify the combustion characteristics further, the mean OH layer location is extracted and compared with experimental measurements. The experimental results are presented with error bars representing the RMS values as presented in the work of Gamba and Mungal\cite{gamba2015ignition}. The results for the DG($p=2$) simulations on each grid are presented in Figure \ref{fig:oh_loc}. In all cases, the simulation captures the initial formation of OH along the windward shear layer. The grids with elements of 300 $\mathrm{\mu}$m or larger predict the mean shear layer location at Y/D $\approx$ 7, while the 250/150 grid is at Y/D $\approx$ 6. This downward displacement of the OH layer may be attributed to its interaction with the shear layer vortices. Although the mean locations differ, the majority of values fall within the RMS spread.

\begin{figure}[htbp]
	\centering
	\includegraphics[scale =0.99]{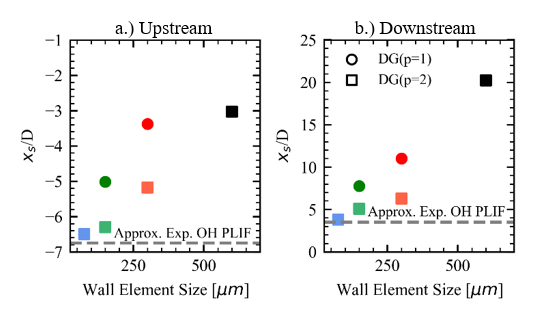}
	\caption{a.) Upstream and b.) downstream flame stabilization points for each polynomial and grid size}
	\label{fig:stabilization}
\end{figure}

\begin{figure}[htbp]
	\centering
	\includegraphics[scale = 0.95]{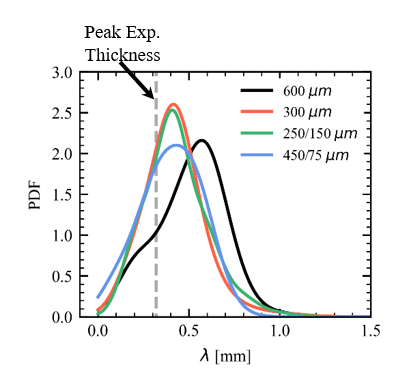}
	\caption{PDFs of OH layer thickness for DG($p=2$) solutions on each grid resolution from -5 < X/D < 5}
	\label{fig:oh_thick}
\end{figure}

\begin{figure*}[htbp]
	\centering
	\includegraphics[scale = 1.0]{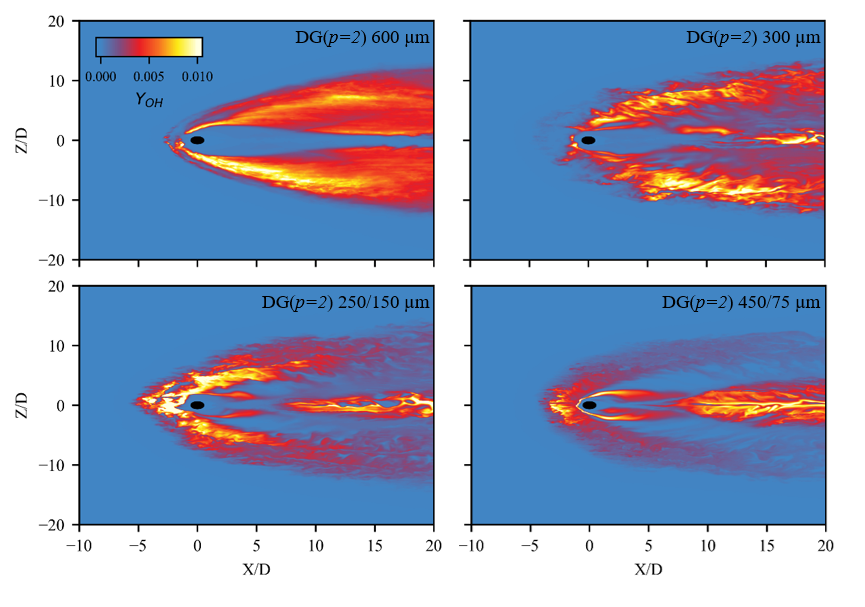}
	\caption{Instantaneous XZ contours of OH mass fraction at Y/D = 0.25 for DG($p=2$) solutions on grids with $h=600$, 300, 250/150, and 450/75 $\mathrm{\mu}$m  at the wall.}
	\label{fig:xz_planes}
\end{figure*}

For further comparisons, the upstream and downstream flame stabilization points relative to the jet injection location are quantified and presented in Figure \ref{fig:stabilization}. Dashed lines indicate the experimental centerplane OH PLIF measurements. The flame stabilization point is defined as the first occurrence of OH along the plate wall as the formation of OH is indicative of heat release. To confirm this, the local heat release rate was extracted and provides similar flame stabilization locations to those based on OH. Examining the upstream flame stabilization first, it is seen that the DG($p=1$) solutions do not provide sufficient resolution to predict the correct flame stabilization point. With DG($p=2$) elements, the flame stabilization location moves closer to the experimental measurements. It is worth noting that when comparing the DG($p=1$) and DG($p=2$) solutions on  the 300 $\mathrm{\mu}$m grids, an increase in polynomial order results in a noticeably more accurate solution as opposed to only a decrease in grid size (despite a similar effective resolution from $h/p+1$). This demonstrates that increasing the polynomial order is an effective approach for obtaining reasonably accurate solutions on coarse, underresolved grids. The refined boundary layer resolution improves the solution accuracy, as shown by the upstream flame stabilization point moving closer to the experimentally observed OH signal.

\begin{figure*}[htbp]
	\centering
	\includegraphics[scale =1.0]{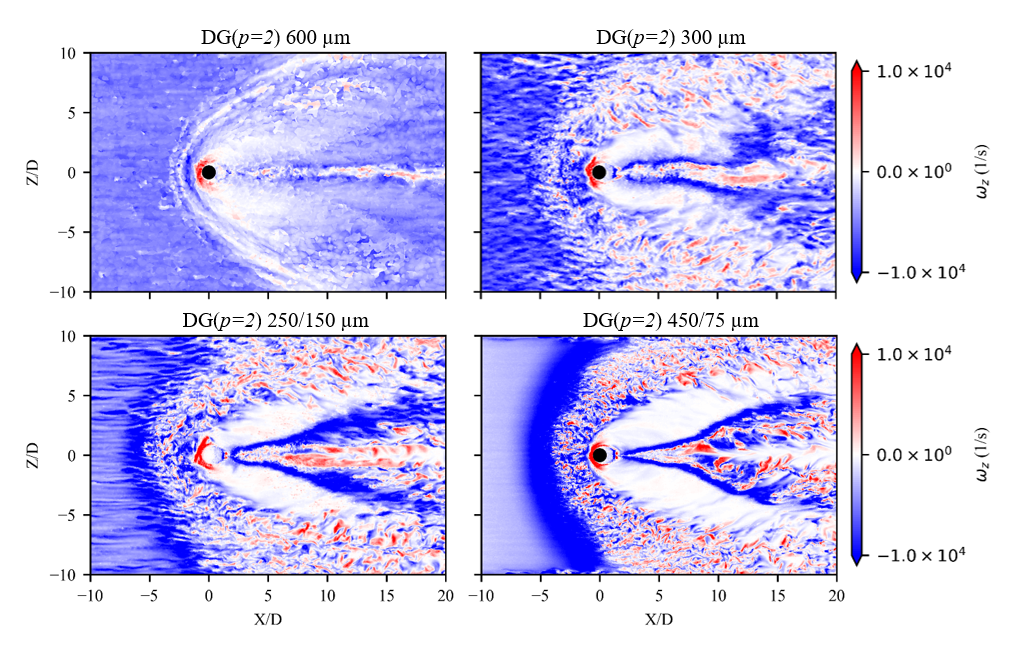}
	\caption{Instantaneous XZ contours of $\omega_z$ at Y/D = 0.25 for DG($p=2$) solutions on grids with $h=600$, 300, 250/150, and 450/75 $\mathrm{\mu}$m.}
	\label{fig:xz_planes_omega}
\end{figure*}
Similar trends are found when examining the downstream flame stabilization location in Figure \ref{fig:stabilization}b. As resolution is increased using both element size and polynomial order, combustion along the plate wall moves closer to the injection point. Once again, as the resolution within the boundary layer is improved, the flame stabilizes closer to the experimental OH PLIF. This is supported by the OH contours presented in Figure \ref{fig:oh_inst}, where it is observed that underresolved solutions are unable to capture immediate mixing and ignition of hydrogen just downstream of the injector.

To further examine the OH signal in the plate centerplane, probability distribution functions (PDF) of the OH layer thickness ($\lambda$) are presented in Figure \ref{fig:oh_thick}. The OH thickness is measured from -5 < X/D < 5 to compare with peak thicknesses measured in the experiment. First the OH mass fraction is isolated using an local thresholding method to isolate the areas where the signal is greater than approximately 5\% of the max \cite{GLASBEY1993532}. The centerline of the structure is determined and then the distance to the edges is then calculated. Experimental measurements show a peak thickness of $\lambda$ = 0.31 mm, while simulations (excluding the 600 $\mathrm{\mu}$m grid) reveal a thickness of approximately $\lambda$ = 0.4 mm. Discrepancies exist between experimental and numerical results. These may arise from differences in sampling size and OH signal thresholding. Furthermore, the chemical reactivity of the hydrogen chemistry model could also contribute to these discrepancies.

Additionally, the experimental results provide OH PLIF images of several XZ planes at different heights above the plate. Instantaneous contours for XZ planes at Y/D = 0.25 for DG($p=2$) solutions on each grid are presented in Figure \ref{fig:xz_planes}. At Y/D = 0.25  are used to quantify the burning along the plate, which extends out to Z/D = $\pm 12.5$ on the 600 $\mathrm{\mu}$m grid and Z/D = $\pm 15$ on the finer grids. This is an underprediction of the spread along the plate compared to the experimental results at Y/D = 0.25 which has OH production expand out to Z/D $\approx$ $\pm 15$.  Additionally, the 600 $\mathrm{\mu}$m mesh has a more constant spread of OH, while the more refined meshes exhibit greater breakup of the OH production, which indicates more turbulent structures being resolved within the boundary layer. The primary differences between the 300 $\mathrm{\mu}$m, 250/150 $\mathrm{\mu}$m and 450/75 meshes is the lower OH mass fraction along the plate.

To further examine the fluid dynamics in the near wall region, instantaneous contours of streamwise vorticity ($\omega_z$) are presented in Figure \ref{fig:xz_planes_omega}. The 600 $\mathrm{\mu}$m mesh, similar to previous results, fails to adequately capture the turbulent flow physics along the plate, resulting in diffuse vortical structures.  Refinement of the near-wall region with smaller elements (250/150 $\mathrm{\mu}$m and 450/75 $\mathrm{\mu}$m) reveals a distinct band of negative vorticity forming where the incoming boundary layer interacts with the separation shock.  This is followed by a highly turbulent region as the flow deflects around the injector. The decrease in OH mass fraction shown in Figure \ref{fig:xz_planes} is attributed to this region. Specifically, increased resolution leads to the formation of more turbulent structures, resulting in higher strain and a more diffuse distribution of OH mass fraction. Downstream of the injector, a strong region of negative vorticity develops around the plate centerline.  Capturing these flow features in greater detail shows a shift in the high-intensity heat release/combustion towards the plate centerline.  Furthermore, increased resolution of the turbulence surrounding the jet correlates with reduced OH production away from the plate centerline, as illustrated in Figure \ref{fig:xz_planes}.

\subsection{Combustion Mode}

\begin{figure}[htbp]
	\centering
	\includegraphics[scale = 0.98]{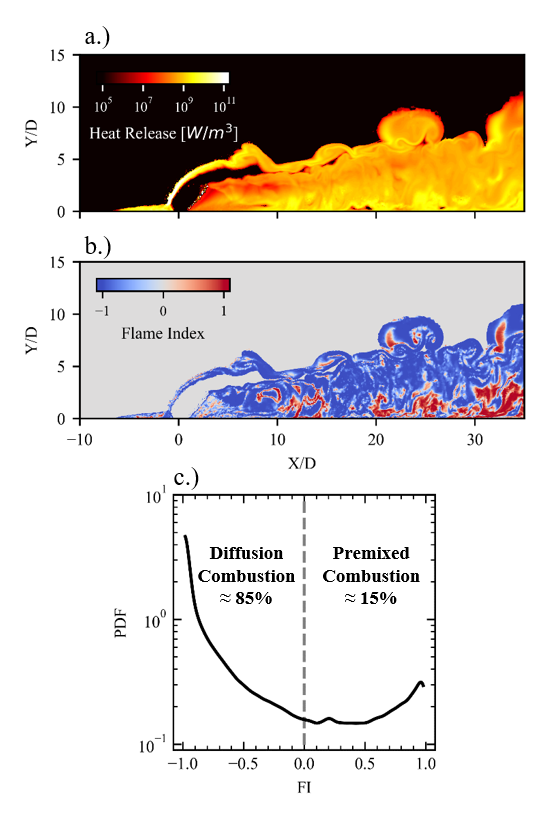}
	\caption{Instantaneous contours of a.) Heat Release Rate and b.) Flame Index filtered by local heat release rate. c.) PDF of flame index}
	\label{fig:FI}
\end{figure}

Flames can be segmented into two modes of combustion, either premixed or diffusion. The flame index is commonly used as a metric to distinguish between locally premixed or diffusion flame in previous studies \cite{yamashita1996numerical,liu2019characteristics}. The flame index (FI) is defined as

\begin{equation}
	\mathrm{FI} = \frac{\nabla Y_{H_2} \cdot \nabla Y_{O_2}}{|\nabla Y_{H_2} \cdot \nabla Y_{O_2}|}.
\end{equation}
In a premixed flame, the gradients between fuel and oxidizer are aligned so that the consumption of reactants across the flame results in a positive flame index. In contrast, the gradients are opposed to each other in a diffusion flame, which is represented by a negative flame index value. The instantaneous heat release and flame index along the plate centerline are presented in Figures \ref{fig:FI}a and \ref{fig:FI}b, respectively. In order to limit the statistics to reacting regions, the flame index is filtered by heat release rate to exclude non-reacting regions. The results show that the combustion is predominantly in a non-premixed combustion mode as the fuel and air cannot sufficiently mix in the relativity short distance downstream from the injector. The premixed combustion primarily resides along the plate within the boundary layer region, while a minor amount resides within the windward shear layer. This is likely due to the fuel distribution presented in Figure \ref{fig:H2}, where there is less fuel near the walls, allowing for mixing to occur more quickly than in the shear layer. The PDF of the flame index is presented in Figure \ref{fig:FI}c to quantify the dominance of the diffusion combustion mode. The data can be constructed into a CDF, which reveals an approximately 85/15\% split between the diffusion and premixed combustion mode, which is consistent with previous research by Liu et. al \cite{liu2019characteristics}.

In the high-enthalpy environment described in this simulation, autoignition is likely to be the dominant mode responsible for the stabilization of the turbulent diffusion flame. This can be quantified using the Damkohler number ($\mathrm{Da}$ or $\mathrm{Da}_I$), which is defined as \cite{yoo2009three,liu2019characteristics}
{
\begin{equation}
	\mathrm{Da} = \frac{|\dot{\omega}_{H_2O}|}{|-\nabla\cdot(\rho Y_{H_2O} V_{\alpha, H_2O})|}
\end{equation}
}

\noindent where $\dot{\omega}_{H_2O}$ is the mass production rate of $H_2O$ and $V_{\alpha, H_2O}$ is the diffusive velocity of $H_2O$. $H_2O$ is selected to quantify the Damkohler number as the production of $H_2O$ in an environment with minimal dissipative losses provides evidence of ignition \cite{yoo2009three,liu2019characteristics}.

\begin{figure}[htbp]
	\centering
	\includegraphics[scale = 1.0]{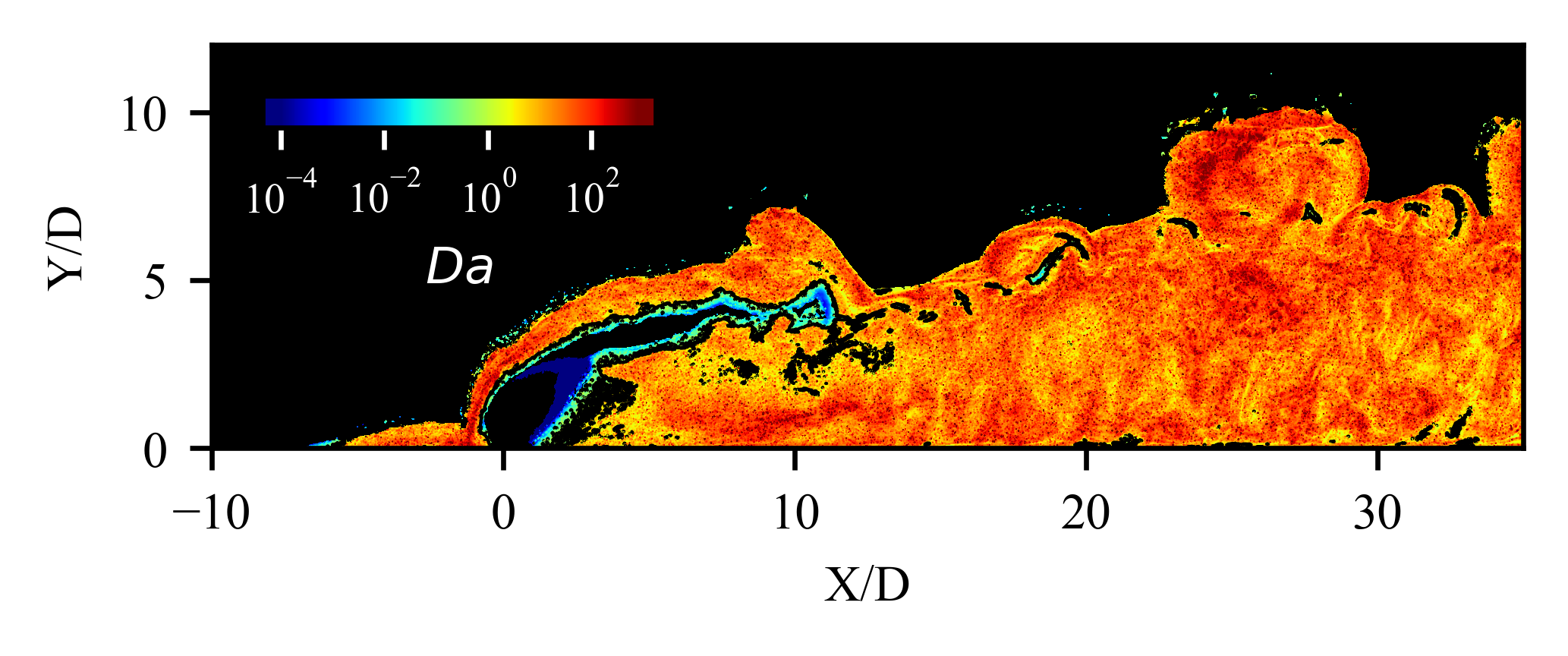}
	\caption{Instantaneous Damkohler number distribution filtered by the local heat release rate along the centerplane.}
	\label{fig:Da}
\end{figure}

An instantaneous $\mathrm{Da}$ distribution along the centerplane is shown in Figure \ref{fig:Da}. The Damkohler number is similarly filtered by the local heat release rate to focus on the reacting regions. The contour reveals that the Damkohler distribution is predominantly larger than 1, which confirms that autoignition is the driving mechanism for the flame stabilization. There are three regions where large ($\mathrm{Da} \gg 1$) Damkohler numbers occur: the bow shock region, along the wall, and in segments of the windward jet shear layer. The black contour isoline is used to illustrate the region where $\mathrm{Da}=1$. The region where $\mathrm{Da}\ll1$ is located near the jet injection point and along the core of the fuel jet trajectory. This is consistent with previous finite volume simulations, although there is a higher contribution of values of $\mathrm{Da}\gg1$ in the downstream region \cite{liu2019characteristics}. A likely cause of this is the hydrogen chemistry mechanism used, which has been shown in previous non-premixed combustion simulations to be over reactive \cite{potturi2014hybrid,li2023application}. Future exploration of available high-fidelity hydrogen chemistry models should be conducted to determine their impact on the statistics described in this section.

\section{Conclusions}
Calculations using a high-order discontinuous Galerkin method on unstructured tetrahedral grids were performed to simulate a reacting hydrogen jet in supersonic crossflow to identify the influence of element size and polynomial order on the solution. Tetrahedral meshes were generated with near-wall element sizes ranging from 600 to 75 $\mathrm{\mu}$m and up to a $p=2$ polynomial approximation (which corresponds to nominally third-order accuracy in space in smooth regions of the flow). The results were compared with the available experimental data to evaluate if underresolved grids using these methods are able to capture the global flame physics. Based on the results the following conclusions can be drawn:
\\
\\
1.) It was shown that increasing from DG($p=0$) to DG($p=2$) elements, the simulations are able capture the expected unsteady physics in the jet shear layers on the same coarse grid illustrating a potential advantage of p-refinement. 
\\
\\
2.) It was also shown that smaller mesh elements were required along the wall to reasonably resolve the separation shock and upstream recirculation zone caused by the jet. However, with DG($p=2$) elements the separation shock was found to agree reasonably with the experimental results while using relatively coarse $h=150$ $\mathrm{\mu}$m elements.
\\
\\
3.) The upstream and downstream flame stabilization locations were also compared across grids and found to be heavily dependent on element size and polynomial order. On coarser grids, p-refinement is more beneficial than h-refinement at capturing the flame stabilization location.
\\
\\
4.) When considering the OH layer thickness it was demonstrated that h-refinement offers minimal improvement on capturing the chemically reacting flow features. DG(p=2) elements with sizes ranging from 450 to 250 $\mathrm{\mu}$m produce similar results demonstrating the capability of $p$-refinement at capturing relevant features which are not directly driven by boundary layer effects. 
\\

This work has shown the applicability of using unstructured tetrahedral meshes with high-order discontinuous Galerkin methods to accurately capture the global flame physics of a reacting jet in supersonic crossflow on underresolved grids. Specifically, key physical traits such as the upstream and downstream flame stabilization points, OH layer thickness, and OH layer location show reasonable agreement with the experimental results of Gamba and Mungal. This agreement suggests that the turbulent mixing behavior of the jet is adequately predicted, even on a comparatively coarse unstructured grid. Future work will explore local (as opposed to global) $p$-adaptivity, investigate different injection schemes, and further examine the coupling of chemistry with these complex fluid dynamics to determine if further grid refinement is necessary.

\begin{acknowledgments}
The authors wish to acknowledge Dr. Eric Marineau of the Hypersonic Aerothermodynamics, High-Speed Propulsion and Materials Program of the Office of Naval Research Code 35 for directly supporting this work.
\end{acknowledgments}

\appendix

\section{Stagnation Pressure and Temperature Boundary Condition\label{sec:StagnationCondition}}

Given an interior static pressure $P^{+}$, solve for the static temperature at the (exterior) boundary, $T^-$, that satisfies

{
\begin{align}
R \ln\left(\frac{P^{+}}{P_{t}}\right)+\sum_{i=1}^{n_{s}}X_{i}s_{i}\left(T_{t}\right)=\sum_{i=1}^{n_{s}}X_{i}s_{i}\left(T^{-}\right),
\end{align}
}
where $R=8314.4621\,\mathrm{J K mol}^{-1}\mathrm{K}^{-1}$ is the universal gas constant,
$P_t$ and $T_t$ are the prescribed stagnation pressure and temperature, respectively,
$n_s$ is the number of species,
$X_i$ is the prescribed mole fraction of species $i$,
and $s_i$ is the specific entropy of species $i$.
The species concentrations at the boundary are then obtained as
\begin{align}
C_{i}^-=\frac{P^{+}}{RT^-}X_{i},
\end{align}
which yields the density at the boundary
\begin{align}
\rho^-=\sum_{i=1}^{n_{s}}W_{i}C_{i}^-
\end{align}
and the mass fractions
\begin{align}
Y_{i}^-=\frac{W_{i}C_{i}^-}{\rho^-}.
\end{align}
Finally, the velocity magnitude at the boundary can be computed using the following  relationship:
{
\begin{align}
\overline{v}^-=\sqrt{2\left(\sum_{i=1}^{n_{s}}Y_{i}^-u_{i}\left(T_{t}\right)-\sum_{i=1}^{n_{s}}Y_{i}^-u_{i}\left(T^-\right)\right)},
\end{align}
where $u_i$ are the internal energies.}
Assuming the flow to be normal to the stagnation vessel, the boundary velocity is then calculated as
\begin{align}
v^-=\overline{v}^-\cdot n
\end{align}
where $n$ is the unit inward surface normal.
The complete boundary state can then be obtained. 

\section*{Author Declarations}

\subsection*{Conflict of Interest}

\noindent The authors have no conflicts to disclose

\subsection*{Author Contributions}
\noindent \textbf{Cal Rising}: Conceptualization (Equal); Methodology (Equal); Investigation (Lead); Formal Analysis (Lead); Visualization (Lead); Writing – original draft (Lead);  Writing – review \& editing (equal). \textbf{Eric Ching}: Conceptualization (Equal); Formal Analysis (Supporting); Methodology (equal); Software (Equal); Writing – original draft (Supporting); Writing – review \& editing (Equal). \textbf{Ryan Johnson}: Conceptualization (Equal); Methodology (Equal); Formal Analysis (Supporting); Software (Equal); Resources (Lead); Funding Acquisition (Lead); Writing - original draft (Supporting); Writing – review \& editing (Equal).

\section*{Data Availability Statement}
\noindent The data that support the findings of this study are available from the corresponding author upon reasonable request subject to the Naval Research Laboratory public release process.

\section*{References}
	\bibliography{apssamp}

\begin{thebibliography}{86}%
\makeatletter
\providecommand \@ifxundefined [1]{%
 \@ifx{#1\undefined}
}%
\providecommand \@ifnum [1]{%
 \ifnum #1\expandafter \@firstoftwo
 \else \expandafter \@secondoftwo
 \fi
}%
\providecommand \@ifx [1]{%
 \ifx #1\expandafter \@firstoftwo
 \else \expandafter \@secondoftwo
 \fi
}%
\providecommand \natexlab [1]{#1}%
\providecommand \enquote  [1]{``#1''}%
\providecommand \bibnamefont  [1]{#1}%
\providecommand \bibfnamefont [1]{#1}%
\providecommand \citenamefont [1]{#1}%
\providecommand \href@noop [0]{\@secondoftwo}%
\providecommand \href [0]{\begingroup \@sanitize@url \@href}%
\providecommand \@href[1]{\@@startlink{#1}\@@href}%
\providecommand \@@href[1]{\endgroup#1\@@endlink}%
\providecommand \@sanitize@url [0]{\catcode `\\12\catcode `\$12\catcode
  `\&12\catcode `\#12\catcode `\^12\catcode `\_12\catcode `\%12\relax}%
\providecommand \@@startlink[1]{}%
\providecommand \@@endlink[0]{}%
\providecommand \url  [0]{\begingroup\@sanitize@url \@url }%
\providecommand \@url [1]{\endgroup\@href {#1}{\urlprefix }}%
\providecommand \urlprefix  [0]{URL }%
\providecommand \Eprint [0]{\href }%
\providecommand \doibase [0]{http://dx.doi.org/}%
\providecommand \selectlanguage [0]{\@gobble}%
\providecommand \bibinfo  [0]{\@secondoftwo}%
\providecommand \bibfield  [0]{\@secondoftwo}%
\providecommand \translation [1]{[#1]}%
\providecommand \BibitemOpen [0]{}%
\providecommand \bibitemStop [0]{}%
\providecommand \bibitemNoStop [0]{.\EOS\space}%
\providecommand \EOS [0]{\spacefactor3000\relax}%
\providecommand \BibitemShut  [1]{\csname bibitem#1\endcsname}%
\let\auto@bib@innerbib\@empty
\bibitem [{\citenamefont {Segal}(2009)}]{segal2009scramjet}%
  \BibitemOpen
  \bibfield  {author} {\bibinfo {author} {\bibfnamefont {C.}~\bibnamefont
  {Segal}},\ }\href@noop {} {\emph {\bibinfo {title} {The scramjet engine:
  processes and characteristics}}},\ Vol.~\bibinfo {volume} {25}\ (\bibinfo
  {publisher} {Cambridge University Press},\ \bibinfo {year}
  {2009})\BibitemShut {NoStop}%
\bibitem [{\citenamefont {Gruber}\ \emph {et~al.}(2004)\citenamefont {Gruber},
  \citenamefont {Donbar}, \citenamefont {Carter},\ and\ \citenamefont
  {Hsu}}]{gruber2004mixing}%
  \BibitemOpen
  \bibfield  {author} {\bibinfo {author} {\bibfnamefont {M.~R.}\ \bibnamefont
  {Gruber}}, \bibinfo {author} {\bibfnamefont {J.~M.}\ \bibnamefont {Donbar}},
  \bibinfo {author} {\bibfnamefont {C.~D.}\ \bibnamefont {Carter}}, \ and\
  \bibinfo {author} {\bibfnamefont {K.-Y.}\ \bibnamefont {Hsu}},\ }\href@noop
  {} {\bibfield  {journal} {\bibinfo  {journal} {Journal of Propulsion and
  Power}\ }\textbf {\bibinfo {volume} {20}},\ \bibinfo {pages} {769} (\bibinfo
  {year} {2004})}\BibitemShut {NoStop}%
\bibitem [{\citenamefont {G{\'e}nin}\ and\ \citenamefont
  {Menon}(2010)}]{genin2010simulation}%
  \BibitemOpen
  \bibfield  {author} {\bibinfo {author} {\bibfnamefont {F.}~\bibnamefont
  {G{\'e}nin}}\ and\ \bibinfo {author} {\bibfnamefont {S.}~\bibnamefont
  {Menon}},\ }\href@noop {} {\bibfield  {journal} {\bibinfo  {journal} {AIAA
  journal}\ }\textbf {\bibinfo {volume} {48}},\ \bibinfo {pages} {526}
  (\bibinfo {year} {2010})}\BibitemShut {NoStop}%
\bibitem [{\citenamefont {F{\"o}rster}\ \emph {et~al.}(2016)\citenamefont
  {F{\"o}rster}, \citenamefont {Dr{\"o}ske}, \citenamefont {B{\"u}hler},
  \citenamefont {von Wolfersdorf},\ and\ \citenamefont
  {Weigand}}]{forster2016analysis}%
  \BibitemOpen
  \bibfield  {author} {\bibinfo {author} {\bibfnamefont {F.~J.}\ \bibnamefont
  {F{\"o}rster}}, \bibinfo {author} {\bibfnamefont {N.~C.}\ \bibnamefont
  {Dr{\"o}ske}}, \bibinfo {author} {\bibfnamefont {M.~N.}\ \bibnamefont
  {B{\"u}hler}}, \bibinfo {author} {\bibfnamefont {J.}~\bibnamefont {von
  Wolfersdorf}}, \ and\ \bibinfo {author} {\bibfnamefont {B.}~\bibnamefont
  {Weigand}},\ }\href@noop {} {\bibfield  {journal} {\bibinfo  {journal}
  {Combustion and Flame}\ }\textbf {\bibinfo {volume} {168}},\ \bibinfo {pages}
  {204} (\bibinfo {year} {2016})}\BibitemShut {NoStop}%
\bibitem [{\citenamefont {Doster}\ \emph {et~al.}(2009)\citenamefont {Doster},
  \citenamefont {King}, \citenamefont {Gruber}, \citenamefont {Carter},
  \citenamefont {Ryan},\ and\ \citenamefont {Hsu}}]{doster2009stream}%
  \BibitemOpen
  \bibfield  {author} {\bibinfo {author} {\bibfnamefont {J.~C.}\ \bibnamefont
  {Doster}}, \bibinfo {author} {\bibfnamefont {P.~I.}\ \bibnamefont {King}},
  \bibinfo {author} {\bibfnamefont {M.~R.}\ \bibnamefont {Gruber}}, \bibinfo
  {author} {\bibfnamefont {C.~D.}\ \bibnamefont {Carter}}, \bibinfo {author}
  {\bibfnamefont {M.~D.}\ \bibnamefont {Ryan}}, \ and\ \bibinfo {author}
  {\bibfnamefont {K.-Y.}\ \bibnamefont {Hsu}},\ }\href@noop {} {\bibfield
  {journal} {\bibinfo  {journal} {Journal of Propulsion and Power}\ }\textbf
  {\bibinfo {volume} {25}},\ \bibinfo {pages} {885} (\bibinfo {year}
  {2009})}\BibitemShut {NoStop}%
\bibitem [{\citenamefont {Vergine}\ \emph {et~al.}(2015)\citenamefont
  {Vergine}, \citenamefont {Crisanti}, \citenamefont {Maddalena}, \citenamefont
  {Miller},\ and\ \citenamefont {Gamba}}]{vergine2015supersonic}%
  \BibitemOpen
  \bibfield  {author} {\bibinfo {author} {\bibfnamefont {F.}~\bibnamefont
  {Vergine}}, \bibinfo {author} {\bibfnamefont {M.}~\bibnamefont {Crisanti}},
  \bibinfo {author} {\bibfnamefont {L.}~\bibnamefont {Maddalena}}, \bibinfo
  {author} {\bibfnamefont {V.}~\bibnamefont {Miller}}, \ and\ \bibinfo {author}
  {\bibfnamefont {M.}~\bibnamefont {Gamba}},\ }\href@noop {} {\bibfield
  {journal} {\bibinfo  {journal} {Journal of Propulsion and Power}\ }\textbf
  {\bibinfo {volume} {31}},\ \bibinfo {pages} {89} (\bibinfo {year}
  {2015})}\BibitemShut {NoStop}%
\bibitem [{\citenamefont {Ben-Yakar}\ and\ \citenamefont
  {Hanson}(2001)}]{ben2001cavity}%
  \BibitemOpen
  \bibfield  {author} {\bibinfo {author} {\bibfnamefont {A.}~\bibnamefont
  {Ben-Yakar}}\ and\ \bibinfo {author} {\bibfnamefont {R.~K.}\ \bibnamefont
  {Hanson}},\ }\href@noop {} {\bibfield  {journal} {\bibinfo  {journal}
  {Journal of propulsion and power}\ }\textbf {\bibinfo {volume} {17}},\
  \bibinfo {pages} {869} (\bibinfo {year} {2001})}\BibitemShut {NoStop}%
\bibitem [{\citenamefont {Gruber}\ \emph {et~al.}(2001)\citenamefont {Gruber},
  \citenamefont {Baurle}, \citenamefont {Mathur},\ and\ \citenamefont
  {Hsu}}]{gruber2001fundamental}%
  \BibitemOpen
  \bibfield  {author} {\bibinfo {author} {\bibfnamefont {M.}~\bibnamefont
  {Gruber}}, \bibinfo {author} {\bibfnamefont {R.}~\bibnamefont {Baurle}},
  \bibinfo {author} {\bibfnamefont {T.}~\bibnamefont {Mathur}}, \ and\ \bibinfo
  {author} {\bibfnamefont {K.-Y.}\ \bibnamefont {Hsu}},\ }\href@noop {}
  {\bibfield  {journal} {\bibinfo  {journal} {Journal of Propulsion and power}\
  }\textbf {\bibinfo {volume} {17}},\ \bibinfo {pages} {146} (\bibinfo {year}
  {2001})}\BibitemShut {NoStop}%
\bibitem [{\citenamefont {Ben-Yakar}\ \emph {et~al.}(2006)\citenamefont
  {Ben-Yakar}, \citenamefont {Mungal},\ and\ \citenamefont
  {Hanson}}]{ben2006time}%
  \BibitemOpen
  \bibfield  {author} {\bibinfo {author} {\bibfnamefont {A.}~\bibnamefont
  {Ben-Yakar}}, \bibinfo {author} {\bibfnamefont {M.}~\bibnamefont {Mungal}}, \
  and\ \bibinfo {author} {\bibfnamefont {R.}~\bibnamefont {Hanson}},\
  }\href@noop {} {\bibfield  {journal} {\bibinfo  {journal} {Physics of
  Fluids}\ }\textbf {\bibinfo {volume} {18}},\ \bibinfo {pages} {026101}
  (\bibinfo {year} {2006})}\BibitemShut {NoStop}%
\bibitem [{\citenamefont {Gamba}\ and\ \citenamefont
  {Mungal}(2015)}]{gamba2015ignition}%
  \BibitemOpen
  \bibfield  {author} {\bibinfo {author} {\bibfnamefont {M.}~\bibnamefont
  {Gamba}}\ and\ \bibinfo {author} {\bibfnamefont {M.~G.}\ \bibnamefont
  {Mungal}},\ }\href@noop {} {\bibfield  {journal} {\bibinfo  {journal}
  {Journal of Fluid Mechanics}\ }\textbf {\bibinfo {volume} {780}},\ \bibinfo
  {pages} {226} (\bibinfo {year} {2015})}\BibitemShut {NoStop}%
\bibitem [{\citenamefont {Sharma}\ \emph {et~al.}(2020)\citenamefont {Sharma},
  \citenamefont {Eswaran},\ and\ \citenamefont
  {Chakraborty}}]{sharma2020effect}%
  \BibitemOpen
  \bibfield  {author} {\bibinfo {author} {\bibfnamefont {V.}~\bibnamefont
  {Sharma}}, \bibinfo {author} {\bibfnamefont {V.}~\bibnamefont {Eswaran}}, \
  and\ \bibinfo {author} {\bibfnamefont {D.}~\bibnamefont {Chakraborty}},\
  }\href@noop {} {\bibfield  {journal} {\bibinfo  {journal} {Aerospace Science
  and Technology}\ }\textbf {\bibinfo {volume} {100}},\ \bibinfo {pages}
  {105786} (\bibinfo {year} {2020})}\BibitemShut {NoStop}%
\bibitem [{\citenamefont {Huang}(2016)}]{huang2016transverse}%
  \BibitemOpen
  \bibfield  {author} {\bibinfo {author} {\bibfnamefont {W.}~\bibnamefont
  {Huang}},\ }\href@noop {} {\bibfield  {journal} {\bibinfo  {journal}
  {Aerospace Science and Technology}\ }\textbf {\bibinfo {volume} {50}},\
  \bibinfo {pages} {183} (\bibinfo {year} {2016})}\BibitemShut {NoStop}%
\bibitem [{\citenamefont {Candler}\ \emph {et~al.}(2017)\citenamefont
  {Candler}, \citenamefont {Cymbalist},\ and\ \citenamefont
  {Dimotakis}}]{candler2017wall}%
  \BibitemOpen
  \bibfield  {author} {\bibinfo {author} {\bibfnamefont {G.~V.}\ \bibnamefont
  {Candler}}, \bibinfo {author} {\bibfnamefont {N.}~\bibnamefont {Cymbalist}},
  \ and\ \bibinfo {author} {\bibfnamefont {P.~E.}\ \bibnamefont {Dimotakis}},\
  }\href@noop {} {\bibfield  {journal} {\bibinfo  {journal} {Aiaa Journal}\
  }\textbf {\bibinfo {volume} {55}},\ \bibinfo {pages} {2410} (\bibinfo {year}
  {2017})}\BibitemShut {NoStop}%
\bibitem [{\citenamefont {Zhao}\ \emph {et~al.}(2017)\citenamefont {Zhao},
  \citenamefont {Zhou}, \citenamefont {Ye}, \citenamefont {Zhu},\ and\
  \citenamefont {Zhang}}]{zhao2017large}%
  \BibitemOpen
  \bibfield  {author} {\bibinfo {author} {\bibfnamefont {M.}~\bibnamefont
  {Zhao}}, \bibinfo {author} {\bibfnamefont {T.}~\bibnamefont {Zhou}}, \bibinfo
  {author} {\bibfnamefont {T.}~\bibnamefont {Ye}}, \bibinfo {author}
  {\bibfnamefont {M.}~\bibnamefont {Zhu}}, \ and\ \bibinfo {author}
  {\bibfnamefont {H.}~\bibnamefont {Zhang}},\ }\href@noop {} {\bibfield
  {journal} {\bibinfo  {journal} {International Journal of Hydrogen Energy}\
  }\textbf {\bibinfo {volume} {42}},\ \bibinfo {pages} {16782} (\bibinfo {year}
  {2017})}\BibitemShut {NoStop}%
\bibitem [{\citenamefont {Nilsson}\ \emph {et~al.}(2021)\citenamefont
  {Nilsson}, \citenamefont {Zhong},\ and\ \citenamefont
  {Fureby}}]{nilsson2021h2}%
  \BibitemOpen
  \bibfield  {author} {\bibinfo {author} {\bibfnamefont {T.}~\bibnamefont
  {Nilsson}}, \bibinfo {author} {\bibfnamefont {S.}~\bibnamefont {Zhong}}, \
  and\ \bibinfo {author} {\bibfnamefont {C.}~\bibnamefont {Fureby}},\
  }\href@noop {} {\bibfield  {journal} {\bibinfo  {journal} {Physics of
  Fluids}\ }\textbf {\bibinfo {volume} {33}} (\bibinfo {year}
  {2021})}\BibitemShut {NoStop}%
\bibitem [{\citenamefont {Saghafian}\ \emph {et~al.}(2015)\citenamefont
  {Saghafian}, \citenamefont {Terrapon},\ and\ \citenamefont
  {Pitsch}}]{saghafian2015efficient}%
  \BibitemOpen
  \bibfield  {author} {\bibinfo {author} {\bibfnamefont {A.}~\bibnamefont
  {Saghafian}}, \bibinfo {author} {\bibfnamefont {V.~E.}\ \bibnamefont
  {Terrapon}}, \ and\ \bibinfo {author} {\bibfnamefont {H.}~\bibnamefont
  {Pitsch}},\ }\href@noop {} {\bibfield  {journal} {\bibinfo  {journal}
  {Combustion and Flame}\ }\textbf {\bibinfo {volume} {162}},\ \bibinfo {pages}
  {652} (\bibinfo {year} {2015})}\BibitemShut {NoStop}%
\bibitem [{\citenamefont {Baurle}(2017)}]{baurle2017hybrid}%
  \BibitemOpen
  \bibfield  {author} {\bibinfo {author} {\bibfnamefont {R.~A.}\ \bibnamefont
  {Baurle}},\ }\href@noop {} {\bibfield  {journal} {\bibinfo  {journal} {AIAA
  Journal}\ }\textbf {\bibinfo {volume} {55}},\ \bibinfo {pages} {544}
  (\bibinfo {year} {2017})}\BibitemShut {NoStop}%
\bibitem [{\citenamefont {Slotnick}\ \emph {et~al.}(2014)\citenamefont
  {Slotnick}, \citenamefont {Khodadoust}, \citenamefont {Alonso}, \citenamefont
  {Darmofal}, \citenamefont {Gropp}, \citenamefont {Lurie},\ and\ \citenamefont
  {Mavriplis}}]{slotnick2014cfd}%
  \BibitemOpen
  \bibfield  {author} {\bibinfo {author} {\bibfnamefont {J.~P.}\ \bibnamefont
  {Slotnick}}, \bibinfo {author} {\bibfnamefont {A.}~\bibnamefont
  {Khodadoust}}, \bibinfo {author} {\bibfnamefont {J.}~\bibnamefont {Alonso}},
  \bibinfo {author} {\bibfnamefont {D.}~\bibnamefont {Darmofal}}, \bibinfo
  {author} {\bibfnamefont {W.}~\bibnamefont {Gropp}}, \bibinfo {author}
  {\bibfnamefont {E.}~\bibnamefont {Lurie}}, \ and\ \bibinfo {author}
  {\bibfnamefont {D.~J.}\ \bibnamefont {Mavriplis}},\ }\href@noop {} {\emph
  {\bibinfo {title} {CFD vision 2030 study: a path to revolutionary
  computational aerosciences}}},\ \bibinfo {type} {Tech. Rep.}\ (\bibinfo
  {year} {2014})\BibitemShut {NoStop}%
\bibitem [{\citenamefont {Peterson}\ and\ \citenamefont
  {Candler}(2011)}]{peterson2011simulations}%
  \BibitemOpen
  \bibfield  {author} {\bibinfo {author} {\bibfnamefont {D.~M.}\ \bibnamefont
  {Peterson}}\ and\ \bibinfo {author} {\bibfnamefont {G.~V.}\ \bibnamefont
  {Candler}},\ }\href@noop {} {\bibfield  {journal} {\bibinfo  {journal} {AIAA
  journal}\ }\textbf {\bibinfo {volume} {49}},\ \bibinfo {pages} {2792}
  (\bibinfo {year} {2011})}\BibitemShut {NoStop}%
\bibitem [{\citenamefont {Hassan}\ \emph {et~al.}(2017)\citenamefont {Hassan},
  \citenamefont {Luke}, \citenamefont {Walters}, \citenamefont {Peterson},
  \citenamefont {Eklund},\ and\ \citenamefont
  {Hagenmaier}}]{hassan2017computations}%
  \BibitemOpen
  \bibfield  {author} {\bibinfo {author} {\bibfnamefont {E.}~\bibnamefont
  {Hassan}}, \bibinfo {author} {\bibfnamefont {E.~A.}\ \bibnamefont {Luke}},
  \bibinfo {author} {\bibfnamefont {D.~K.}\ \bibnamefont {Walters}}, \bibinfo
  {author} {\bibfnamefont {D.~M.}\ \bibnamefont {Peterson}}, \bibinfo {author}
  {\bibfnamefont {D.}~\bibnamefont {Eklund}}, \ and\ \bibinfo {author}
  {\bibfnamefont {M.}~\bibnamefont {Hagenmaier}},\ }\href@noop {} {\bibfield
  {journal} {\bibinfo  {journal} {Flow, Turbulence and Combustion}\ }\textbf
  {\bibinfo {volume} {99}},\ \bibinfo {pages} {437} (\bibinfo {year}
  {2017})}\BibitemShut {NoStop}%
\bibitem [{\citenamefont {Nielsen}\ \emph {et~al.}(2021)\citenamefont
  {Nielsen}, \citenamefont {Edwards}, \citenamefont {Chelliah}, \citenamefont
  {Lieber}, \citenamefont {Geipel}, \citenamefont {Goyne}, \citenamefont
  {Rockwell},\ and\ \citenamefont {Cutler}}]{nielsen2021hybrid}%
  \BibitemOpen
  \bibfield  {author} {\bibinfo {author} {\bibfnamefont {T.~B.}\ \bibnamefont
  {Nielsen}}, \bibinfo {author} {\bibfnamefont {J.~R.}\ \bibnamefont
  {Edwards}}, \bibinfo {author} {\bibfnamefont {H.~K.}\ \bibnamefont
  {Chelliah}}, \bibinfo {author} {\bibfnamefont {D.}~\bibnamefont {Lieber}},
  \bibinfo {author} {\bibfnamefont {C.}~\bibnamefont {Geipel}}, \bibinfo
  {author} {\bibfnamefont {C.~P.}\ \bibnamefont {Goyne}}, \bibinfo {author}
  {\bibfnamefont {R.~D.}\ \bibnamefont {Rockwell}}, \ and\ \bibinfo {author}
  {\bibfnamefont {A.~D.}\ \bibnamefont {Cutler}},\ }\href@noop {} {\bibfield
  {journal} {\bibinfo  {journal} {AIAA Journal}\ }\textbf {\bibinfo {volume}
  {59}},\ \bibinfo {pages} {2440} (\bibinfo {year} {2021})}\BibitemShut
  {NoStop}%
\bibitem [{\citenamefont {Boles}\ \emph {et~al.}(2010)\citenamefont {Boles},
  \citenamefont {Edwards},\ and\ \citenamefont {Baurle}}]{boles2010large}%
  \BibitemOpen
  \bibfield  {author} {\bibinfo {author} {\bibfnamefont {J.~A.}\ \bibnamefont
  {Boles}}, \bibinfo {author} {\bibfnamefont {J.~R.}\ \bibnamefont {Edwards}},
  \ and\ \bibinfo {author} {\bibfnamefont {R.~A.}\ \bibnamefont {Baurle}},\
  }\href@noop {} {\bibfield  {journal} {\bibinfo  {journal} {AIAA journal}\
  }\textbf {\bibinfo {volume} {48}},\ \bibinfo {pages} {1444} (\bibinfo {year}
  {2010})}\BibitemShut {NoStop}%
\bibitem [{\citenamefont {Peterson}\ and\ \citenamefont
  {Hassan}(2018)}]{peterson2018hybrid}%
  \BibitemOpen
  \bibfield  {author} {\bibinfo {author} {\bibfnamefont {D.~M.}\ \bibnamefont
  {Peterson}}\ and\ \bibinfo {author} {\bibfnamefont {E.~A.}\ \bibnamefont
  {Hassan}},\ }in\ \href@noop {} {\emph {\bibinfo {booktitle} {2018 AIAA
  Aerospace Sciences Meeting}}}\ (\bibinfo {year} {2018})\ p.\ \bibinfo {pages}
  {1144}\BibitemShut {NoStop}%
\bibitem [{\citenamefont {Peterson}(2023)}]{peterson2023simulation}%
  \BibitemOpen
  \bibfield  {author} {\bibinfo {author} {\bibfnamefont {D.~M.}\ \bibnamefont
  {Peterson}},\ }\href@noop {} {\bibfield  {journal} {\bibinfo  {journal}
  {Proceedings of the Combustion Institute}\ }\textbf {\bibinfo {volume}
  {39}},\ \bibinfo {pages} {3137} (\bibinfo {year} {2023})}\BibitemShut
  {NoStop}%
\bibitem [{\citenamefont {Gassner}\ and\ \citenamefont
  {Beck}(2013)}]{gassner2013accuracy}%
  \BibitemOpen
  \bibfield  {author} {\bibinfo {author} {\bibfnamefont {G.~J.}\ \bibnamefont
  {Gassner}}\ and\ \bibinfo {author} {\bibfnamefont {A.~D.}\ \bibnamefont
  {Beck}},\ }\href@noop {} {\bibfield  {journal} {\bibinfo  {journal}
  {Theoretical and Computational Fluid Dynamics}\ }\textbf {\bibinfo {volume}
  {27}},\ \bibinfo {pages} {221} (\bibinfo {year} {2013})}\BibitemShut
  {NoStop}%
\bibitem [{\citenamefont {Beck}\ \emph {et~al.}(2014)\citenamefont {Beck},
  \citenamefont {Bolemann}, \citenamefont {Flad}, \citenamefont {Frank},
  \citenamefont {Gassner}, \citenamefont {Hindenlang},\ and\ \citenamefont
  {Munz}}]{beck2014high}%
  \BibitemOpen
  \bibfield  {author} {\bibinfo {author} {\bibfnamefont {A.~D.}\ \bibnamefont
  {Beck}}, \bibinfo {author} {\bibfnamefont {T.}~\bibnamefont {Bolemann}},
  \bibinfo {author} {\bibfnamefont {D.}~\bibnamefont {Flad}}, \bibinfo {author}
  {\bibfnamefont {H.}~\bibnamefont {Frank}}, \bibinfo {author} {\bibfnamefont
  {G.~J.}\ \bibnamefont {Gassner}}, \bibinfo {author} {\bibfnamefont
  {F.}~\bibnamefont {Hindenlang}}, \ and\ \bibinfo {author} {\bibfnamefont
  {C.-D.}\ \bibnamefont {Munz}},\ }\href@noop {} {\bibfield  {journal}
  {\bibinfo  {journal} {International Journal for Numerical Methods in Fluids}\
  }\textbf {\bibinfo {volume} {76}},\ \bibinfo {pages} {522} (\bibinfo {year}
  {2014})}\BibitemShut {NoStop}%
\bibitem [{\citenamefont {Fernandez}\ \emph {et~al.}(2017)\citenamefont
  {Fernandez}, \citenamefont {Nguyen},\ and\ \citenamefont
  {Peraire}}]{fernandez2017hybridized}%
  \BibitemOpen
  \bibfield  {author} {\bibinfo {author} {\bibfnamefont {P.}~\bibnamefont
  {Fernandez}}, \bibinfo {author} {\bibfnamefont {N.~C.}\ \bibnamefont
  {Nguyen}}, \ and\ \bibinfo {author} {\bibfnamefont {J.}~\bibnamefont
  {Peraire}},\ }\href@noop {} {\bibfield  {journal} {\bibinfo  {journal}
  {Journal of Computational Physics}\ }\textbf {\bibinfo {volume} {336}},\
  \bibinfo {pages} {308} (\bibinfo {year} {2017})}\BibitemShut {NoStop}%
\bibitem [{\citenamefont {Lv}\ and\ \citenamefont
  {Ekaterinaris}(2023)}]{lv2023recent}%
  \BibitemOpen
  \bibfield  {author} {\bibinfo {author} {\bibfnamefont {Y.}~\bibnamefont
  {Lv}}\ and\ \bibinfo {author} {\bibfnamefont {J.}~\bibnamefont
  {Ekaterinaris}},\ }\href@noop {} {\bibfield  {journal} {\bibinfo  {journal}
  {Progress in Aerospace Sciences}\ ,\ \bibinfo {pages} {100929}} (\bibinfo
  {year} {2023})}\BibitemShut {NoStop}%
\bibitem [{\citenamefont {Hoskin}\ \emph {et~al.}(2024)\citenamefont {Hoskin},
  \citenamefont {Van~Heyningen}, \citenamefont {Nguyen}, \citenamefont
  {Vila-P{\'e}rez}, \citenamefont {Harris},\ and\ \citenamefont
  {Peraire}}]{hoskin2024discontinuous}%
  \BibitemOpen
  \bibfield  {author} {\bibinfo {author} {\bibfnamefont {D.~S.}\ \bibnamefont
  {Hoskin}}, \bibinfo {author} {\bibfnamefont {R.~L.}\ \bibnamefont
  {Van~Heyningen}}, \bibinfo {author} {\bibfnamefont {N.~C.}\ \bibnamefont
  {Nguyen}}, \bibinfo {author} {\bibfnamefont {J.}~\bibnamefont
  {Vila-P{\'e}rez}}, \bibinfo {author} {\bibfnamefont {W.~L.}\ \bibnamefont
  {Harris}}, \ and\ \bibinfo {author} {\bibfnamefont {J.}~\bibnamefont
  {Peraire}},\ }\href@noop {} {\bibfield  {journal} {\bibinfo  {journal}
  {Progress in Aerospace Sciences}\ }\textbf {\bibinfo {volume} {146}},\
  \bibinfo {pages} {100999} (\bibinfo {year} {2024})}\BibitemShut {NoStop}%
\bibitem [{\citenamefont {Ceze}\ and\ \citenamefont
  {Fidkowski}(2013)}]{ceze2013anisotropic}%
  \BibitemOpen
  \bibfield  {author} {\bibinfo {author} {\bibfnamefont {M.}~\bibnamefont
  {Ceze}}\ and\ \bibinfo {author} {\bibfnamefont {K.~J.}\ \bibnamefont
  {Fidkowski}},\ }\href@noop {} {\bibfield  {journal} {\bibinfo  {journal}
  {AIAA journal}\ }\textbf {\bibinfo {volume} {51}},\ \bibinfo {pages} {492}
  (\bibinfo {year} {2013})}\BibitemShut {NoStop}%
\bibitem [{\citenamefont {Froehle}\ and\ \citenamefont
  {Persson}(2014)}]{froehle2014high}%
  \BibitemOpen
  \bibfield  {author} {\bibinfo {author} {\bibfnamefont {B.}~\bibnamefont
  {Froehle}}\ and\ \bibinfo {author} {\bibfnamefont {P.-O.}\ \bibnamefont
  {Persson}},\ }\href@noop {} {\bibfield  {journal} {\bibinfo  {journal}
  {Journal of Computational Physics}\ }\textbf {\bibinfo {volume} {272}},\
  \bibinfo {pages} {455} (\bibinfo {year} {2014})}\BibitemShut {NoStop}%
\bibitem [{\citenamefont {Arndt}\ \emph {et~al.}(2020)\citenamefont {Arndt},
  \citenamefont {Fehn}, \citenamefont {Kanschat}, \citenamefont {Kormann},
  \citenamefont {Kronbichler}, \citenamefont {Munch}, \citenamefont {Wall},\
  and\ \citenamefont {Witte}}]{arndt2020exadg}%
  \BibitemOpen
  \bibfield  {author} {\bibinfo {author} {\bibfnamefont {D.}~\bibnamefont
  {Arndt}}, \bibinfo {author} {\bibfnamefont {N.}~\bibnamefont {Fehn}},
  \bibinfo {author} {\bibfnamefont {G.}~\bibnamefont {Kanschat}}, \bibinfo
  {author} {\bibfnamefont {K.}~\bibnamefont {Kormann}}, \bibinfo {author}
  {\bibfnamefont {M.}~\bibnamefont {Kronbichler}}, \bibinfo {author}
  {\bibfnamefont {P.}~\bibnamefont {Munch}}, \bibinfo {author} {\bibfnamefont
  {W.~A.}\ \bibnamefont {Wall}}, \ and\ \bibinfo {author} {\bibfnamefont
  {J.}~\bibnamefont {Witte}},\ }in\ \href@noop {} {\emph {\bibinfo {booktitle}
  {Software for exascale computing-SPPEXA 2016-2019}}}\ (\bibinfo
  {organization} {Springer International Publishing},\ \bibinfo {year} {2020})\
  pp.\ \bibinfo {pages} {189--224}\BibitemShut {NoStop}%
\bibitem [{\citenamefont {Cuong~Nguyen}\ \emph {et~al.}(2022)\citenamefont
  {Cuong~Nguyen}, \citenamefont {Terrana},\ and\ \citenamefont
  {Peraire}}]{cuong2022large}%
  \BibitemOpen
  \bibfield  {author} {\bibinfo {author} {\bibfnamefont {N.}~\bibnamefont
  {Cuong~Nguyen}}, \bibinfo {author} {\bibfnamefont {S.}~\bibnamefont
  {Terrana}}, \ and\ \bibinfo {author} {\bibfnamefont {J.}~\bibnamefont
  {Peraire}},\ }\href@noop {} {\bibfield  {journal} {\bibinfo  {journal} {AIAA
  Journal}\ }\textbf {\bibinfo {volume} {60}},\ \bibinfo {pages} {3060}
  (\bibinfo {year} {2022})}\BibitemShut {NoStop}%
\bibitem [{\citenamefont {Ching}\ \emph
  {et~al.}(2024{\natexlab{a}})\citenamefont {Ching}, \citenamefont {Johnson},\
  and\ \citenamefont {Kercher}}]{Chi22}%
  \BibitemOpen
  \bibfield  {author} {\bibinfo {author} {\bibfnamefont {E.~J.}\ \bibnamefont
  {Ching}}, \bibinfo {author} {\bibfnamefont {R.~F.}\ \bibnamefont {Johnson}},
  \ and\ \bibinfo {author} {\bibfnamefont {A.~D.}\ \bibnamefont {Kercher}},\
  }\href@noop {} {\bibfield  {journal} {\bibinfo  {journal} {Journal of
  Computational Physics}\ ,\ \bibinfo {pages} {112881}} (\bibinfo {year}
  {2024}{\natexlab{a}})}\BibitemShut {NoStop}%
\bibitem [{\citenamefont {Ching}\ \emph
  {et~al.}(2024{\natexlab{b}})\citenamefont {Ching}, \citenamefont {Johnson},\
  and\ \citenamefont {Kercher}}]{Chi22_2}%
  \BibitemOpen
  \bibfield  {author} {\bibinfo {author} {\bibfnamefont {E.~J.}\ \bibnamefont
  {Ching}}, \bibinfo {author} {\bibfnamefont {R.~F.}\ \bibnamefont {Johnson}},
  \ and\ \bibinfo {author} {\bibfnamefont {A.~D.}\ \bibnamefont {Kercher}},\
  }\href@noop {} {\bibfield  {journal} {\bibinfo  {journal} {Journal of
  Computational Physics}\ ,\ \bibinfo {pages} {112878}} (\bibinfo {year}
  {2024}{\natexlab{b}})}\BibitemShut {NoStop}%
\bibitem [{\citenamefont {Ching}\ \emph {et~al.}(2025)\citenamefont {Ching},
  \citenamefont {Johnson}, \citenamefont {Burrows}, \citenamefont {Higgs},\
  and\ \citenamefont {Kercher}}]{Chi25}%
  \BibitemOpen
  \bibfield  {author} {\bibinfo {author} {\bibfnamefont {E.~J.}\ \bibnamefont
  {Ching}}, \bibinfo {author} {\bibfnamefont {R.~F.}\ \bibnamefont {Johnson}},
  \bibinfo {author} {\bibfnamefont {S.}~\bibnamefont {Burrows}}, \bibinfo
  {author} {\bibfnamefont {J.}~\bibnamefont {Higgs}}, \ and\ \bibinfo {author}
  {\bibfnamefont {A.~D.}\ \bibnamefont {Kercher}},\ }\href@noop {} {\bibfield
  {journal} {\bibinfo  {journal} {Journal of Computational Physics}\ }\textbf
  {\bibinfo {volume} {525}},\ \bibinfo {pages} {113746} (\bibinfo {year}
  {2025})}\BibitemShut {NoStop}%
\bibitem [{\citenamefont {Liu}\ \emph {et~al.}(2019)\citenamefont {Liu},
  \citenamefont {Yu}, \citenamefont {Wang}, \citenamefont {Sun}, \citenamefont
  {Wang},\ and\ \citenamefont {Grosshans}}]{liu2019characteristics}%
  \BibitemOpen
  \bibfield  {author} {\bibinfo {author} {\bibfnamefont {C.}~\bibnamefont
  {Liu}}, \bibinfo {author} {\bibfnamefont {J.}~\bibnamefont {Yu}}, \bibinfo
  {author} {\bibfnamefont {Z.}~\bibnamefont {Wang}}, \bibinfo {author}
  {\bibfnamefont {M.}~\bibnamefont {Sun}}, \bibinfo {author} {\bibfnamefont
  {H.}~\bibnamefont {Wang}}, \ and\ \bibinfo {author} {\bibfnamefont
  {H.}~\bibnamefont {Grosshans}},\ }\href@noop {} {\bibfield  {journal}
  {\bibinfo  {journal} {Physics of Fluids}\ }\textbf {\bibinfo {volume} {31}}
  (\bibinfo {year} {2019})}\BibitemShut {NoStop}%
\bibitem [{\citenamefont {Lv}\ \emph {et~al.}(2018)\citenamefont {Lv},
  \citenamefont {Ma},\ and\ \citenamefont {Ihme}}]{lv2018underresolved}%
  \BibitemOpen
  \bibfield  {author} {\bibinfo {author} {\bibfnamefont {Y.}~\bibnamefont
  {Lv}}, \bibinfo {author} {\bibfnamefont {P.~C.}\ \bibnamefont {Ma}}, \ and\
  \bibinfo {author} {\bibfnamefont {M.}~\bibnamefont {Ihme}},\ }\href@noop {}
  {\bibfield  {journal} {\bibinfo  {journal} {Computers \& Fluids}\ }\textbf
  {\bibinfo {volume} {161}},\ \bibinfo {pages} {89} (\bibinfo {year}
  {2018})}\BibitemShut {NoStop}%
\bibitem [{\citenamefont {Spalart}(2000)}]{spalart2000strategies}%
  \BibitemOpen
  \bibfield  {author} {\bibinfo {author} {\bibfnamefont {P.~R.}\ \bibnamefont
  {Spalart}},\ }\href@noop {} {\bibfield  {journal} {\bibinfo  {journal}
  {International journal of heat and fluid flow}\ }\textbf {\bibinfo {volume}
  {21}},\ \bibinfo {pages} {252} (\bibinfo {year} {2000})}\BibitemShut
  {NoStop}%
\bibitem [{\citenamefont {Edwards}\ \emph {et~al.}(2024)\citenamefont
  {Edwards}, \citenamefont {Rajath},\ and\ \citenamefont
  {Navratil}}]{edwards2024multi}%
  \BibitemOpen
  \bibfield  {author} {\bibinfo {author} {\bibfnamefont {J.~R.}\ \bibnamefont
  {Edwards}}, \bibinfo {author} {\bibfnamefont {C.}~\bibnamefont {Rajath}}, \
  and\ \bibinfo {author} {\bibfnamefont {A.~V.}\ \bibnamefont {Navratil}},\
  }\href@noop {} {\bibfield  {journal} {\bibinfo  {journal} {Computers \&
  Fluids}\ }\textbf {\bibinfo {volume} {275}},\ \bibinfo {pages} {106235}
  (\bibinfo {year} {2024})}\BibitemShut {NoStop}%
\bibitem [{\citenamefont {P{\'e}quin}\ \emph {et~al.}(2022)\citenamefont
  {P{\'e}quin}, \citenamefont {Iavarone}, \citenamefont {Malpica~Galassi},\
  and\ \citenamefont {Parente}}]{pequin2022partially}%
  \BibitemOpen
  \bibfield  {author} {\bibinfo {author} {\bibfnamefont {A.}~\bibnamefont
  {P{\'e}quin}}, \bibinfo {author} {\bibfnamefont {S.}~\bibnamefont
  {Iavarone}}, \bibinfo {author} {\bibfnamefont {R.}~\bibnamefont
  {Malpica~Galassi}}, \ and\ \bibinfo {author} {\bibfnamefont {A.}~\bibnamefont
  {Parente}},\ }\href@noop {} {\bibfield  {journal} {\bibinfo  {journal}
  {Physics of Fluids}\ }\textbf {\bibinfo {volume} {34}} (\bibinfo {year}
  {2022})}\BibitemShut {NoStop}%
\bibitem [{\citenamefont {Johnson}\ and\ \citenamefont
  {Kercher}(2020)}]{johnson2020conservative}%
  \BibitemOpen
  \bibfield  {author} {\bibinfo {author} {\bibfnamefont {R.~F.}\ \bibnamefont
  {Johnson}}\ and\ \bibinfo {author} {\bibfnamefont {A.~D.}\ \bibnamefont
  {Kercher}},\ }\href@noop {} {\bibfield  {journal} {\bibinfo  {journal}
  {Journal of Computational Physics}\ }\textbf {\bibinfo {volume} {423}},\
  \bibinfo {pages} {109826} (\bibinfo {year} {2020})}\BibitemShut {NoStop}%
\bibitem [{\citenamefont {Lv}\ and\ \citenamefont
  {Ihme}(2014)}]{lv2014discontinuous}%
  \BibitemOpen
  \bibfield  {author} {\bibinfo {author} {\bibfnamefont {Y.}~\bibnamefont
  {Lv}}\ and\ \bibinfo {author} {\bibfnamefont {M.}~\bibnamefont {Ihme}},\
  }\href@noop {} {\bibfield  {journal} {\bibinfo  {journal} {Journal of
  Computational Physics}\ }\textbf {\bibinfo {volume} {270}},\ \bibinfo {pages}
  {105} (\bibinfo {year} {2014})}\BibitemShut {NoStop}%
\bibitem [{\citenamefont {Lv}\ and\ \citenamefont {Ihme}(2017)}]{lv2017high}%
  \BibitemOpen
  \bibfield  {author} {\bibinfo {author} {\bibfnamefont {Y.}~\bibnamefont
  {Lv}}\ and\ \bibinfo {author} {\bibfnamefont {M.}~\bibnamefont {Ihme}},\
  }\href@noop {} {\bibfield  {journal} {\bibinfo  {journal} {Acta Mechanica
  Sinica}\ }\textbf {\bibinfo {volume} {33}},\ \bibinfo {pages} {486} (\bibinfo
  {year} {2017})}\BibitemShut {NoStop}%
\bibitem [{\citenamefont {Guti{\'e}rrez-Jorquera}\ and\ \citenamefont
  {Kummer}(2022)}]{gutierrez2022fully}%
  \BibitemOpen
  \bibfield  {author} {\bibinfo {author} {\bibfnamefont {J.}~\bibnamefont
  {Guti{\'e}rrez-Jorquera}}\ and\ \bibinfo {author} {\bibfnamefont
  {F.}~\bibnamefont {Kummer}},\ }\href@noop {} {\bibfield  {journal} {\bibinfo
  {journal} {International Journal for Numerical Methods in Fluids}\ }\textbf
  {\bibinfo {volume} {94}},\ \bibinfo {pages} {316} (\bibinfo {year}
  {2022})}\BibitemShut {NoStop}%
\bibitem [{\citenamefont {Rising}\ \emph {et~al.}(2022)\citenamefont {Rising},
  \citenamefont {Goodwin}, \citenamefont {Johnson}, \citenamefont {Kessler},
  \citenamefont {Sosa}, \citenamefont {Thornton},\ and\ \citenamefont
  {Ahmed}}]{rising2022numerical}%
  \BibitemOpen
  \bibfield  {author} {\bibinfo {author} {\bibfnamefont {C.~J.}\ \bibnamefont
  {Rising}}, \bibinfo {author} {\bibfnamefont {G.~B.}\ \bibnamefont {Goodwin}},
  \bibinfo {author} {\bibfnamefont {R.~F.}\ \bibnamefont {Johnson}}, \bibinfo
  {author} {\bibfnamefont {D.~A.}\ \bibnamefont {Kessler}}, \bibinfo {author}
  {\bibfnamefont {J.}~\bibnamefont {Sosa}}, \bibinfo {author} {\bibfnamefont
  {M.}~\bibnamefont {Thornton}}, \ and\ \bibinfo {author} {\bibfnamefont
  {K.~A.}\ \bibnamefont {Ahmed}},\ }\href@noop {} {\bibfield  {journal}
  {\bibinfo  {journal} {Aerospace Science and Technology}\ }\textbf {\bibinfo
  {volume} {129}},\ \bibinfo {pages} {107805} (\bibinfo {year}
  {2022})}\BibitemShut {NoStop}%
\bibitem [{\citenamefont {Terrana}\ \emph {et~al.}(2020)\citenamefont
  {Terrana}, \citenamefont {Nguyen},\ and\ \citenamefont
  {Peraire}}]{terrana2020gpu}%
  \BibitemOpen
  \bibfield  {author} {\bibinfo {author} {\bibfnamefont {S.}~\bibnamefont
  {Terrana}}, \bibinfo {author} {\bibfnamefont {C.}~\bibnamefont {Nguyen}}, \
  and\ \bibinfo {author} {\bibfnamefont {J.}~\bibnamefont {Peraire}},\ }in\
  \href@noop {} {\emph {\bibinfo {booktitle} {AIAA Scitech 2020 Forum}}}\
  (\bibinfo {year} {2020})\ p.\ \bibinfo {pages} {1062}\BibitemShut {NoStop}%
\bibitem [{\citenamefont {Ching}\ \emph
  {et~al.}(2019{\natexlab{a}})\citenamefont {Ching}, \citenamefont {Lv},
  \citenamefont {Gnoffo}, \citenamefont {Barnhardt},\ and\ \citenamefont
  {Ihme}}]{ching2019shock}%
  \BibitemOpen
  \bibfield  {author} {\bibinfo {author} {\bibfnamefont {E.~J.}\ \bibnamefont
  {Ching}}, \bibinfo {author} {\bibfnamefont {Y.}~\bibnamefont {Lv}}, \bibinfo
  {author} {\bibfnamefont {P.}~\bibnamefont {Gnoffo}}, \bibinfo {author}
  {\bibfnamefont {M.}~\bibnamefont {Barnhardt}}, \ and\ \bibinfo {author}
  {\bibfnamefont {M.}~\bibnamefont {Ihme}},\ }\href@noop {} {\bibfield
  {journal} {\bibinfo  {journal} {Journal of Computational Physics}\ }\textbf
  {\bibinfo {volume} {376}},\ \bibinfo {pages} {54} (\bibinfo {year}
  {2019}{\natexlab{a}})}\BibitemShut {NoStop}%
\bibitem [{\citenamefont {Bai}\ and\ \citenamefont
  {Fidkowski}(2022)}]{bai2022continuous}%
  \BibitemOpen
  \bibfield  {author} {\bibinfo {author} {\bibfnamefont {Y.}~\bibnamefont
  {Bai}}\ and\ \bibinfo {author} {\bibfnamefont {K.~J.}\ \bibnamefont
  {Fidkowski}},\ }\href@noop {} {\bibfield  {journal} {\bibinfo  {journal}
  {AIAA Journal}\ }\textbf {\bibinfo {volume} {60}},\ \bibinfo {pages} {5678}
  (\bibinfo {year} {2022})}\BibitemShut {NoStop}%
\bibitem [{\citenamefont {Barter}\ and\ \citenamefont
  {Darmofal}(2010)}]{barter2010shock}%
  \BibitemOpen
  \bibfield  {author} {\bibinfo {author} {\bibfnamefont {G.~E.}\ \bibnamefont
  {Barter}}\ and\ \bibinfo {author} {\bibfnamefont {D.~L.}\ \bibnamefont
  {Darmofal}},\ }\href@noop {} {\bibfield  {journal} {\bibinfo  {journal}
  {Journal of Computational Physics}\ }\textbf {\bibinfo {volume} {229}},\
  \bibinfo {pages} {1810} (\bibinfo {year} {2010})}\BibitemShut {NoStop}%
\bibitem [{\citenamefont {Moxey}\ \emph {et~al.}(2017)\citenamefont {Moxey},
  \citenamefont {Cantwell}, \citenamefont {Mengaldo}, \citenamefont {Serson},
  \citenamefont {Ekelschot}, \citenamefont {Peir{\'o}}, \citenamefont
  {Sherwin},\ and\ \citenamefont {Kirby}}]{moxey2017towards}%
  \BibitemOpen
  \bibfield  {author} {\bibinfo {author} {\bibfnamefont {D.}~\bibnamefont
  {Moxey}}, \bibinfo {author} {\bibfnamefont {C.}~\bibnamefont {Cantwell}},
  \bibinfo {author} {\bibfnamefont {G.}~\bibnamefont {Mengaldo}}, \bibinfo
  {author} {\bibfnamefont {D.}~\bibnamefont {Serson}}, \bibinfo {author}
  {\bibfnamefont {D.}~\bibnamefont {Ekelschot}}, \bibinfo {author}
  {\bibfnamefont {J.}~\bibnamefont {Peir{\'o}}}, \bibinfo {author}
  {\bibfnamefont {S.}~\bibnamefont {Sherwin}}, \ and\ \bibinfo {author}
  {\bibfnamefont {R.}~\bibnamefont {Kirby}},\ }in\ \href@noop {} {\emph
  {\bibinfo {booktitle} {Spectral and High Order Methods for Partial
  Differential Equations ICOSAHOM 2016: Selected Papers from the ICOSAHOM
  conference, June 27-July 1, 2016, Rio de Janeiro, Brazil}}}\ (\bibinfo
  {organization} {Springer},\ \bibinfo {year} {2017})\ pp.\ \bibinfo {pages}
  {63--79}\BibitemShut {NoStop}%
\bibitem [{\citenamefont {Wang}\ \emph {et~al.}(2013)\citenamefont {Wang},
  \citenamefont {Fidkowski}, \citenamefont {Abgrall}, \citenamefont {Bassi},
  \citenamefont {Caraeni}, \citenamefont {Cary}, \citenamefont {Deconinck},
  \citenamefont {Hartmann}, \citenamefont {Hillewaert}, \citenamefont {Huynh}
  \emph {et~al.}}]{wang2013high}%
  \BibitemOpen
  \bibfield  {author} {\bibinfo {author} {\bibfnamefont {Z.~J.}\ \bibnamefont
  {Wang}}, \bibinfo {author} {\bibfnamefont {K.}~\bibnamefont {Fidkowski}},
  \bibinfo {author} {\bibfnamefont {R.}~\bibnamefont {Abgrall}}, \bibinfo
  {author} {\bibfnamefont {F.}~\bibnamefont {Bassi}}, \bibinfo {author}
  {\bibfnamefont {D.}~\bibnamefont {Caraeni}}, \bibinfo {author} {\bibfnamefont
  {A.}~\bibnamefont {Cary}}, \bibinfo {author} {\bibfnamefont {H.}~\bibnamefont
  {Deconinck}}, \bibinfo {author} {\bibfnamefont {R.}~\bibnamefont {Hartmann}},
  \bibinfo {author} {\bibfnamefont {K.}~\bibnamefont {Hillewaert}}, \bibinfo
  {author} {\bibfnamefont {H.~T.}\ \bibnamefont {Huynh}},  \emph {et~al.},\
  }\href@noop {} {\bibfield  {journal} {\bibinfo  {journal} {International
  Journal for Numerical Methods in Fluids}\ }\textbf {\bibinfo {volume} {72}},\
  \bibinfo {pages} {811} (\bibinfo {year} {2013})}\BibitemShut {NoStop}%
\bibitem [{\citenamefont {Naddei}\ \emph {et~al.}(2019)\citenamefont {Naddei},
  \citenamefont {de~la Llave~Plata}, \citenamefont {Couaillier},\ and\
  \citenamefont {Coquel}}]{naddei2019comparison}%
  \BibitemOpen
  \bibfield  {author} {\bibinfo {author} {\bibfnamefont {F.}~\bibnamefont
  {Naddei}}, \bibinfo {author} {\bibfnamefont {M.}~\bibnamefont {de~la
  Llave~Plata}}, \bibinfo {author} {\bibfnamefont {V.}~\bibnamefont
  {Couaillier}}, \ and\ \bibinfo {author} {\bibfnamefont {F.}~\bibnamefont
  {Coquel}},\ }\href@noop {} {\bibfield  {journal} {\bibinfo  {journal}
  {Journal of Computational Physics}\ }\textbf {\bibinfo {volume} {376}},\
  \bibinfo {pages} {508} (\bibinfo {year} {2019})}\BibitemShut {NoStop}%
\bibitem [{\citenamefont {Marchal}\ \emph {et~al.}(2023)\citenamefont
  {Marchal}, \citenamefont {Deniau}, \citenamefont {Boussuge}, \citenamefont
  {Cuenot},\ and\ \citenamefont {Mercier}}]{marchal2023extension}%
  \BibitemOpen
  \bibfield  {author} {\bibinfo {author} {\bibfnamefont {T.}~\bibnamefont
  {Marchal}}, \bibinfo {author} {\bibfnamefont {H.}~\bibnamefont {Deniau}},
  \bibinfo {author} {\bibfnamefont {J.-F.}\ \bibnamefont {Boussuge}}, \bibinfo
  {author} {\bibfnamefont {B.}~\bibnamefont {Cuenot}}, \ and\ \bibinfo {author}
  {\bibfnamefont {R.}~\bibnamefont {Mercier}},\ }\href@noop {} {\bibfield
  {journal} {\bibinfo  {journal} {Flow, Turbulence and Combustion}\ }\textbf
  {\bibinfo {volume} {111}},\ \bibinfo {pages} {141} (\bibinfo {year}
  {2023})}\BibitemShut {NoStop}%
\bibitem [{\citenamefont {Yoon}\ \emph {et~al.}(2007)\citenamefont {Yoon},
  \citenamefont {Gnoffo}, \citenamefont {White},\ and\ \citenamefont
  {Thomas}}]{yoon2007computational}%
  \BibitemOpen
  \bibfield  {author} {\bibinfo {author} {\bibfnamefont {S.}~\bibnamefont
  {Yoon}}, \bibinfo {author} {\bibfnamefont {P.}~\bibnamefont {Gnoffo}},
  \bibinfo {author} {\bibfnamefont {J.}~\bibnamefont {White}}, \ and\ \bibinfo
  {author} {\bibfnamefont {J.}~\bibnamefont {Thomas}},\ }in\ \href@noop {}
  {\emph {\bibinfo {booktitle} {39th AIAA Thermophysics Conference}}}\
  (\bibinfo {year} {2007})\ p.\ \bibinfo {pages} {4265}\BibitemShut {NoStop}%
\bibitem [{\citenamefont {Gnoffo}(2007)}]{gnoffo2007simulation}%
  \BibitemOpen
  \bibfield  {author} {\bibinfo {author} {\bibfnamefont {P.}~\bibnamefont
  {Gnoffo}},\ }in\ \href@noop {} {\emph {\bibinfo {booktitle} {18th AIAA
  computational fluid dynamics conference}}}\ (\bibinfo {year} {2007})\ p.\
  \bibinfo {pages} {3960}\BibitemShut {NoStop}%
\bibitem [{\citenamefont {Candler}\ \emph {et~al.}(2009)\citenamefont
  {Candler}, \citenamefont {Mavriplis},\ and\ \citenamefont
  {Trevino}}]{candler2009current}%
  \BibitemOpen
  \bibfield  {author} {\bibinfo {author} {\bibfnamefont {G.}~\bibnamefont
  {Candler}}, \bibinfo {author} {\bibfnamefont {D.}~\bibnamefont {Mavriplis}},
  \ and\ \bibinfo {author} {\bibfnamefont {L.}~\bibnamefont {Trevino}},\ }in\
  \href@noop {} {\emph {\bibinfo {booktitle} {47th AIAA aerospace sciences
  meeting including the new horizons forum and aerospace exposition}}}\
  (\bibinfo {year} {2009})\ p.\ \bibinfo {pages} {153}\BibitemShut {NoStop}%
\bibitem [{\citenamefont {Candler}(2015)}]{candler2015next}%
  \BibitemOpen
  \bibfield  {author} {\bibinfo {author} {\bibfnamefont {G.~V.}\ \bibnamefont
  {Candler}},\ }in\ \href@noop {} {\emph {\bibinfo {booktitle} {22nd AIAA
  Computational Fluid Dynamics Conference}}}\ (\bibinfo {year} {2015})\ p.\
  \bibinfo {pages} {3048}\BibitemShut {NoStop}%
\bibitem [{\citenamefont {Sharma}\ \emph {et~al.}(2024)\citenamefont {Sharma},
  \citenamefont {Singh}, \citenamefont {Angelilli},\ and\ \citenamefont
  {Raman}}]{sharma2024interaction}%
  \BibitemOpen
  \bibfield  {author} {\bibinfo {author} {\bibfnamefont {S.}~\bibnamefont
  {Sharma}}, \bibinfo {author} {\bibfnamefont {J.}~\bibnamefont {Singh}},
  \bibinfo {author} {\bibfnamefont {L.}~\bibnamefont {Angelilli}}, \ and\
  \bibinfo {author} {\bibfnamefont {V.}~\bibnamefont {Raman}},\ }\href@noop {}
  {\bibfield  {journal} {\bibinfo  {journal} {Proceedings of the Combustion
  Institute}\ }\textbf {\bibinfo {volume} {40}},\ \bibinfo {pages} {105295}
  (\bibinfo {year} {2024})}\BibitemShut {NoStop}%
\bibitem [{\citenamefont {Rising}\ \emph {et~al.}(2024)\citenamefont {Rising},
  \citenamefont {Goodwin}, \citenamefont {Ching}, \citenamefont {Viswanath},\
  and\ \citenamefont {Johnson}}]{rising2024use}%
  \BibitemOpen
  \bibfield  {author} {\bibinfo {author} {\bibfnamefont {C.}~\bibnamefont
  {Rising}}, \bibinfo {author} {\bibfnamefont {G.~B.}\ \bibnamefont {Goodwin}},
  \bibinfo {author} {\bibfnamefont {E.~J.}\ \bibnamefont {Ching}}, \bibinfo
  {author} {\bibfnamefont {K.}~\bibnamefont {Viswanath}}, \ and\ \bibinfo
  {author} {\bibfnamefont {R.}~\bibnamefont {Johnson}},\ }in\ \href@noop {}
  {\emph {\bibinfo {booktitle} {AIAA SCITECH 2024 Forum}}}\ (\bibinfo {year}
  {2024})\ p.\ \bibinfo {pages} {0911}\BibitemShut {NoStop}%
\bibitem [{\citenamefont {Kee}\ \emph {et~al.}(1989{\natexlab{a}})\citenamefont
  {Kee}, \citenamefont {Rupley},\ and\ \citenamefont {Miller}}]{chemkin89}%
  \BibitemOpen
  \bibfield  {author} {\bibinfo {author} {\bibfnamefont {R.}~\bibnamefont
  {Kee}}, \bibinfo {author} {\bibfnamefont {F.}~\bibnamefont {Rupley}}, \ and\
  \bibinfo {author} {\bibfnamefont {J.}~\bibnamefont {Miller}},\ }\href@noop {}
  {\  (\bibinfo {year} {1989}{\natexlab{a}})}\BibitemShut {NoStop}%
\bibitem [{\citenamefont {McBride}\ \emph {et~al.}(1993)\citenamefont
  {McBride}, \citenamefont {Gordon},\ and\ \citenamefont {Reno}}]{Mcb93}%
  \BibitemOpen
  \bibfield  {author} {\bibinfo {author} {\bibfnamefont {B.~J.}\ \bibnamefont
  {McBride}}, \bibinfo {author} {\bibfnamefont {S.}~\bibnamefont {Gordon}}, \
  and\ \bibinfo {author} {\bibfnamefont {M.~A.}\ \bibnamefont {Reno}},\
  }\href@noop {} {\  (\bibinfo {year} {1993})}\BibitemShut {NoStop}%
\bibitem [{\citenamefont {McBride}\ \emph {et~al.}(2002)\citenamefont
  {McBride}, \citenamefont {Zehe},\ and\ \citenamefont {Gordon}}]{Mcb02}%
  \BibitemOpen
  \bibfield  {author} {\bibinfo {author} {\bibfnamefont {B.~J.}\ \bibnamefont
  {McBride}}, \bibinfo {author} {\bibfnamefont {M.~J.}\ \bibnamefont {Zehe}}, \
  and\ \bibinfo {author} {\bibfnamefont {S.}~\bibnamefont {Gordon}},\
  }\href@noop {} {\  (\bibinfo {year} {2002})}\BibitemShut {NoStop}%
\bibitem [{\citenamefont {Coffee}\ and\ \citenamefont
  {Heimerl}(1981)}]{coffee1981transport}%
  \BibitemOpen
  \bibfield  {author} {\bibinfo {author} {\bibfnamefont {T.~P.}\ \bibnamefont
  {Coffee}}\ and\ \bibinfo {author} {\bibfnamefont {J.}~\bibnamefont
  {Heimerl}},\ }\href@noop {} {\bibfield  {journal} {\bibinfo  {journal}
  {Combustion and Flame}\ }\textbf {\bibinfo {volume} {43}},\ \bibinfo {pages}
  {273} (\bibinfo {year} {1981})}\BibitemShut {NoStop}%
\bibitem [{\citenamefont {Houim}\ and\ \citenamefont
  {Kuo}(2011)}]{houim2011low}%
  \BibitemOpen
  \bibfield  {author} {\bibinfo {author} {\bibfnamefont {R.~W.}\ \bibnamefont
  {Houim}}\ and\ \bibinfo {author} {\bibfnamefont {K.~K.}\ \bibnamefont
  {Kuo}},\ }\href@noop {} {\bibfield  {journal} {\bibinfo  {journal} {Journal
  of Computational Physics}\ }\textbf {\bibinfo {volume} {230}},\ \bibinfo
  {pages} {8527} (\bibinfo {year} {2011})}\BibitemShut {NoStop}%
\bibitem [{\citenamefont {Kee}\ \emph {et~al.}(1989{\natexlab{b}})\citenamefont
  {Kee}, \citenamefont {Miller}, \citenamefont {Evans},\ and\ \citenamefont
  {Dixon-Lewis}}]{Kee89}%
  \BibitemOpen
  \bibfield  {author} {\bibinfo {author} {\bibfnamefont {R.~J.}\ \bibnamefont
  {Kee}}, \bibinfo {author} {\bibfnamefont {J.~A.}\ \bibnamefont {Miller}},
  \bibinfo {author} {\bibfnamefont {G.~H.}\ \bibnamefont {Evans}}, \ and\
  \bibinfo {author} {\bibfnamefont {G.}~\bibnamefont {Dixon-Lewis}},\ }\href
  {\doibase https://doi.org/10.1016/S0082-0784(89)80158-4} {\bibfield
  {journal} {\bibinfo  {journal} {Symposium (International) on Combustion}\
  }\textbf {\bibinfo {volume} {22}},\ \bibinfo {pages} {1479 } (\bibinfo {year}
  {1989}{\natexlab{b}})}\BibitemShut {NoStop}%
\bibitem [{\citenamefont {R.~Wilke}(1950)}]{Wil50}%
  \BibitemOpen
  \bibfield  {author} {\bibinfo {author} {\bibfnamefont {C.}~\bibnamefont
  {R.~Wilke}},\ }\href {\doibase 10.1063/1.1747673} {\bibfield  {journal}
  {\bibinfo  {journal} {J.~Chem.~Phys}\ }\textbf {\bibinfo {volume} {18}},\
  \bibinfo {pages} {517} (\bibinfo {year} {1950})}\BibitemShut {NoStop}%
\bibitem [{\citenamefont {{Mathur}}\ \emph {et~al.}(1967)\citenamefont
  {{Mathur}}, \citenamefont {{Tondon}},\ and\ \citenamefont
  {{Saxena}}}]{Mat67}%
  \BibitemOpen
  \bibfield  {author} {\bibinfo {author} {\bibfnamefont {S.}~\bibnamefont
  {{Mathur}}}, \bibinfo {author} {\bibfnamefont {P.~K.}\ \bibnamefont
  {{Tondon}}}, \ and\ \bibinfo {author} {\bibfnamefont {S.~C.}\ \bibnamefont
  {{Saxena}}},\ }\href {\doibase 10.1080/00268976700100731} {\bibfield
  {journal} {\bibinfo  {journal} {Molecular Physics}\ }\textbf {\bibinfo
  {volume} {12}},\ \bibinfo {pages} {569} (\bibinfo {year} {1967})}\BibitemShut
  {NoStop}%
\bibitem [{\citenamefont {Westbrook}(1982)}]{WESTBROOK1982191}%
  \BibitemOpen
  \bibfield  {author} {\bibinfo {author} {\bibfnamefont {C.~K.}\ \bibnamefont
  {Westbrook}},\ }\href@noop {} {\bibfield  {journal} {\bibinfo  {journal}
  {Combustion and Flame}\ }\textbf {\bibinfo {volume} {46}},\ \bibinfo {pages}
  {191} (\bibinfo {year} {1982})}\BibitemShut {NoStop}%
\bibitem [{\citenamefont {Toro}(2013)}]{Tor13}%
  \BibitemOpen
  \bibfield  {author} {\bibinfo {author} {\bibfnamefont {E.}~\bibnamefont
  {Toro}},\ }\href@noop {} {\emph {\bibinfo {title} {Riemann solvers and
  numerical methods for fluid dynamics: {A} practical introduction}}}\
  (\bibinfo  {publisher} {Springer Science \& Business Media},\ \bibinfo {year}
  {2013})\BibitemShut {NoStop}%
\bibitem [{\citenamefont {Bassi}\ and\ \citenamefont {Rebay}(2000)}]{Bas00}%
  \BibitemOpen
  \bibfield  {author} {\bibinfo {author} {\bibfnamefont {F.}~\bibnamefont
  {Bassi}}\ and\ \bibinfo {author} {\bibfnamefont {S.}~\bibnamefont {Rebay}},\
  }in\ \href@noop {} {\emph {\bibinfo {booktitle} {Discontinuous Galerkin
  Methods}}}\ (\bibinfo  {publisher} {Springer},\ \bibinfo {year} {2000})\ pp.\
  \bibinfo {pages} {197--208}\BibitemShut {NoStop}%
\bibitem [{\citenamefont {Gottlieb}\ \emph {et~al.}(2001)\citenamefont
  {Gottlieb}, \citenamefont {Shu},\ and\ \citenamefont {Tadmor}}]{Got01}%
  \BibitemOpen
  \bibfield  {author} {\bibinfo {author} {\bibfnamefont {S.}~\bibnamefont
  {Gottlieb}}, \bibinfo {author} {\bibfnamefont {C.}~\bibnamefont {Shu}}, \
  and\ \bibinfo {author} {\bibfnamefont {E.}~\bibnamefont {Tadmor}},\
  }\href@noop {} {\bibfield  {journal} {\bibinfo  {journal} {SIAM review}\
  }\textbf {\bibinfo {volume} {43}},\ \bibinfo {pages} {89} (\bibinfo {year}
  {2001})}\BibitemShut {NoStop}%
\bibitem [{\citenamefont {Ketcheson}(2008)}]{Ket08}%
  \BibitemOpen
  \bibfield  {author} {\bibinfo {author} {\bibfnamefont {D.~I.}\ \bibnamefont
  {Ketcheson}},\ }\href@noop {} {\bibfield  {journal} {\bibinfo  {journal}
  {SIAM Journal on Scientific Computing}\ }\textbf {\bibinfo {volume} {30}},\
  \bibinfo {pages} {2113} (\bibinfo {year} {2008})}\BibitemShut {NoStop}%
\bibitem [{\citenamefont {Rodriguez}\ \emph {et~al.}(2018)\citenamefont
  {Rodriguez}, \citenamefont {Aftosmis},\ and\ \citenamefont
  {Nemec}}]{rodriguez2018formulation}%
  \BibitemOpen
  \bibfield  {author} {\bibinfo {author} {\bibfnamefont {D.~L.}\ \bibnamefont
  {Rodriguez}}, \bibinfo {author} {\bibfnamefont {M.~J.}\ \bibnamefont
  {Aftosmis}}, \ and\ \bibinfo {author} {\bibfnamefont {M.}~\bibnamefont
  {Nemec}},\ }in\ \href@noop {} {\emph {\bibinfo {booktitle} {2018 AIAA
  Aerospace Sciences Meeting}}}\ (\bibinfo {year} {2018})\ p.\ \bibinfo {pages}
  {0334}\BibitemShut {NoStop}%
\bibitem [{\citenamefont {{Lawrence Livermore National
  Laboratory}}(2023)}]{llnl_lassen}%
  \BibitemOpen
  \bibfield  {author} {\bibinfo {author} {\bibnamefont {{Lawrence Livermore
  National Laboratory}}},\ }\href
  {https://hpc.llnl.gov/hardware/compute-platforms/lassen} {\enquote {\bibinfo
  {title} {Lassen},}\ } (\bibinfo {year} {2023}),\ \bibinfo {note} {accessed:
  2024-08-22}\BibitemShut {NoStop}%
\bibitem [{\citenamefont {Fric}\ and\ \citenamefont
  {Roshko}(1994)}]{fric1994vortical}%
  \BibitemOpen
  \bibfield  {author} {\bibinfo {author} {\bibfnamefont {T.}~\bibnamefont
  {Fric}}\ and\ \bibinfo {author} {\bibfnamefont {A.}~\bibnamefont {Roshko}},\
  }\href@noop {} {\bibfield  {journal} {\bibinfo  {journal} {Journal of Fluid
  Mechanics}\ }\textbf {\bibinfo {volume} {279}},\ \bibinfo {pages} {1}
  (\bibinfo {year} {1994})}\BibitemShut {NoStop}%
\bibitem [{\citenamefont {Dowdy}\ and\ \citenamefont
  {Newton}(1963)}]{dowdy1963investigation}%
  \BibitemOpen
  \bibfield  {author} {\bibinfo {author} {\bibfnamefont {M.~W.}\ \bibnamefont
  {Dowdy}}\ and\ \bibinfo {author} {\bibfnamefont {J.}~\bibnamefont {Newton}},\
  }\href@noop {} {\bibfield  {journal} {\bibinfo  {journal} {Jet Propulsion
  Lab. Technical Report}\ ,\ \bibinfo {pages} {120}} (\bibinfo {year}
  {1963})}\BibitemShut {NoStop}%
\bibitem [{\citenamefont {{Siemens Digital Industries Software}}(2023)}]{ccm0}%
  \BibitemOpen
  \bibfield  {author} {\bibinfo {author} {\bibnamefont {{Siemens Digital
  Industries Software}}},\ }\href@noop {} {\enquote {\bibinfo {title}
  {Simcenter \uppercase{STAR-CCM+}, version 2306},}\ } (\bibinfo {year}
  {Siemens 2023})\BibitemShut {NoStop}%
\bibitem [{\citenamefont {Moura}\ \emph {et~al.}(2017)\citenamefont {Moura},
  \citenamefont {Mengaldo}, \citenamefont {Peir{\'o}},\ and\ \citenamefont
  {Sherwin}}]{moura2017eddy}%
  \BibitemOpen
  \bibfield  {author} {\bibinfo {author} {\bibfnamefont {R.~C.}\ \bibnamefont
  {Moura}}, \bibinfo {author} {\bibfnamefont {G.}~\bibnamefont {Mengaldo}},
  \bibinfo {author} {\bibfnamefont {J.}~\bibnamefont {Peir{\'o}}}, \ and\
  \bibinfo {author} {\bibfnamefont {S.~J.}\ \bibnamefont {Sherwin}},\
  }\href@noop {} {\bibfield  {journal} {\bibinfo  {journal} {Journal of
  Computational Physics}\ }\textbf {\bibinfo {volume} {330}},\ \bibinfo {pages}
  {615} (\bibinfo {year} {2017})}\BibitemShut {NoStop}%
\bibitem [{\citenamefont {Ching}\ \emph
  {et~al.}(2019{\natexlab{b}})\citenamefont {Ching}, \citenamefont {Lv},
  \citenamefont {Gnoffo}, \citenamefont {Barnhardt},\ and\ \citenamefont
  {Ihme}}]{Chi19}%
  \BibitemOpen
  \bibfield  {author} {\bibinfo {author} {\bibfnamefont {E.}~\bibnamefont
  {Ching}}, \bibinfo {author} {\bibfnamefont {Y.}~\bibnamefont {Lv}}, \bibinfo
  {author} {\bibfnamefont {P.}~\bibnamefont {Gnoffo}}, \bibinfo {author}
  {\bibfnamefont {M.}~\bibnamefont {Barnhardt}}, \ and\ \bibinfo {author}
  {\bibfnamefont {M.}~\bibnamefont {Ihme}},\ }\href@noop {} {\bibfield
  {journal} {\bibinfo  {journal} {Journal of Computational Physics}\ }\textbf
  {\bibinfo {volume} {376}},\ \bibinfo {pages} {54} (\bibinfo {year}
  {2019}{\natexlab{b}})}\BibitemShut {NoStop}%
\bibitem [{\citenamefont {Pizzaia}\ and\ \citenamefont
  {Rossmann}(2018)}]{pizzaia2018effect}%
  \BibitemOpen
  \bibfield  {author} {\bibinfo {author} {\bibfnamefont {A.}~\bibnamefont
  {Pizzaia}}\ and\ \bibinfo {author} {\bibfnamefont {T.}~\bibnamefont
  {Rossmann}},\ }\href@noop {} {\bibfield  {journal} {\bibinfo  {journal}
  {Physics of Fluids}\ }\textbf {\bibinfo {volume} {30}} (\bibinfo {year}
  {2018})}\BibitemShut {NoStop}%
\bibitem [{\citenamefont {Glasbey}(1993)}]{GLASBEY1993532}%
  \BibitemOpen
  \bibfield  {author} {\bibinfo {author} {\bibfnamefont {C.}~\bibnamefont
  {Glasbey}},\ }\href {\doibase https://doi.org/10.1006/cgip.1993.1040}
  {\bibfield  {journal} {\bibinfo  {journal} {CVGIP: Graphical Models and Image
  Processing}\ }\textbf {\bibinfo {volume} {55}},\ \bibinfo {pages} {532}
  (\bibinfo {year} {1993})}\BibitemShut {NoStop}%
\bibitem [{\citenamefont {Yamashita}\ \emph {et~al.}(1996)\citenamefont
  {Yamashita}, \citenamefont {Shimada},\ and\ \citenamefont
  {Takeno}}]{yamashita1996numerical}%
  \BibitemOpen
  \bibfield  {author} {\bibinfo {author} {\bibfnamefont {H.}~\bibnamefont
  {Yamashita}}, \bibinfo {author} {\bibfnamefont {M.}~\bibnamefont {Shimada}},
  \ and\ \bibinfo {author} {\bibfnamefont {T.}~\bibnamefont {Takeno}},\ }in\
  \href@noop {} {\emph {\bibinfo {booktitle} {Symposium (international) on
  combustion}}},\ Vol.~\bibinfo {volume} {26}\ (\bibinfo {organization}
  {Elsevier},\ \bibinfo {year} {1996})\ pp.\ \bibinfo {pages}
  {27--34}\BibitemShut {NoStop}%
\bibitem [{\citenamefont {Yoo}\ \emph {et~al.}(2009)\citenamefont {Yoo},
  \citenamefont {Sankaran},\ and\ \citenamefont {Chen}}]{yoo2009three}%
  \BibitemOpen
  \bibfield  {author} {\bibinfo {author} {\bibfnamefont {C.~S.}\ \bibnamefont
  {Yoo}}, \bibinfo {author} {\bibfnamefont {R.}~\bibnamefont {Sankaran}}, \
  and\ \bibinfo {author} {\bibfnamefont {J.}~\bibnamefont {Chen}},\ }\href@noop
  {} {\bibfield  {journal} {\bibinfo  {journal} {Journal of Fluid Mechanics}\
  }\textbf {\bibinfo {volume} {640}},\ \bibinfo {pages} {453} (\bibinfo {year}
  {2009})}\BibitemShut {NoStop}%
\bibitem [{\citenamefont {Potturi}\ and\ \citenamefont
  {Edwards}(2014)}]{potturi2014hybrid}%
  \BibitemOpen
  \bibfield  {author} {\bibinfo {author} {\bibfnamefont {A.~S.}\ \bibnamefont
  {Potturi}}\ and\ \bibinfo {author} {\bibfnamefont {J.~R.}\ \bibnamefont
  {Edwards}},\ }\href@noop {} {\bibfield  {journal} {\bibinfo  {journal} {AIAA
  journal}\ }\textbf {\bibinfo {volume} {52}},\ \bibinfo {pages} {1417}
  (\bibinfo {year} {2014})}\BibitemShut {NoStop}%
\bibitem [{\citenamefont {Li}\ \emph {et~al.}(2023)\citenamefont {Li},
  \citenamefont {Wang},\ and\ \citenamefont {Li}}]{li2023application}%
  \BibitemOpen
  \bibfield  {author} {\bibinfo {author} {\bibfnamefont {Z.}~\bibnamefont
  {Li}}, \bibinfo {author} {\bibfnamefont {J.}~\bibnamefont {Wang}}, \ and\
  \bibinfo {author} {\bibfnamefont {X.}~\bibnamefont {Li}},\ }\href@noop {}
  {\bibfield  {journal} {\bibinfo  {journal} {Fuel}\ }\textbf {\bibinfo
  {volume} {349}},\ \bibinfo {pages} {128659} (\bibinfo {year}
  {2023})}\BibitemShut {NoStop}%
\end{thebibliography}%
\end{document}